\documentclass[twocolumn,numberedappendix,twocolappendix,iop]{openjournal}

\usepackage{xcolor}
\usepackage{multirow}
\usepackage{amsmath}
\usepackage[colorlinks,linkcolor=blue,citecolor=blue,urlcolor=blue]{hyperref}

\begin{document}

\title{Stellar reddening map from DESI imaging and spectroscopy}

\author{\vspace{-1.1cm}
Rongpu~Zhou\altaffilmark{1,*},
Julien~Guy\altaffilmark{1},
Sergey~E.~Koposov\altaffilmark{2,3},
Edward~F.~Schlafly\altaffilmark{4},
David~Schlegel\altaffilmark{1},
Jessica~Aguilar\altaffilmark{1},
Steven~Ahlen\altaffilmark{5},
Stephen~Bailey\altaffilmark{1},
David~Bianchi\altaffilmark{6},
David~Brooks\altaffilmark{7},
Edmond~Chaussidon\altaffilmark{1},
Todd~Claybaugh\altaffilmark{1},
Kyle~Dawson\altaffilmark{8},
Axel~de la Macorra\altaffilmark{9},
Arjun~Dey\altaffilmark{22},
Biprateep~Dey\altaffilmark{10},
Daniel~J.~Eisenstein\altaffilmark{11},
Simone~Ferraro\altaffilmark{1,12},
Andreu~Font-Ribera\altaffilmark{7,13},
Jaime~E.~Forero-Romero\altaffilmark{14,15},
Enrique~Gaztañaga\altaffilmark{16,17,18},
Satya~Gontcho A Gontcho\altaffilmark{1},
Gaston~Gutierrez\altaffilmark{19},
Klaus~Honscheid\altaffilmark{20,21},
Stephanie~Juneau\altaffilmark{22},
Robert~Kehoe\altaffilmark{23},
David~Kirkby\altaffilmark{24},
Theodore~Kisner\altaffilmark{1},
Anthony~Kremin\altaffilmark{1},
Andrew~Lambert\altaffilmark{1},
Martin~Landriau\altaffilmark{1},
Laurent~Le~Guillou\altaffilmark{25},
Michael~E.~Levi\altaffilmark{1},
Ting~S.~Li\altaffilmark{26},
Marc~Manera\altaffilmark{27,13},
Paul~Martini\altaffilmark{20,28},
Aaron~Meisner\altaffilmark{22},
Ramon~Miquel\altaffilmark{29,13},
John~Moustakas\altaffilmark{30},
Adam~D.~Myers\altaffilmark{31},
Jeffrey~ A.~Newman\altaffilmark{10},
Gustavo~Niz\altaffilmark{32,33},
Nathalie~Palanque-Delabrouille\altaffilmark{34,1},
Will~J.~Percival\altaffilmark{35,36,37},
Claire~Poppett\altaffilmark{1,38,12},
Francisco~Prada\altaffilmark{39},
Anand~Raichoor\altaffilmark{1},
Ashley~J.~Ross\altaffilmark{20},
Graziano~Rossi\altaffilmark{40},
Eusebio~Sanchez\altaffilmark{41},
Andrew~K.~Saydjari\altaffilmark{42},
Michael~Schubnell\altaffilmark{43,44},
David~Sprayberry\altaffilmark{22},
Gregory~Tarl\'{e}\altaffilmark{44},
Benjamin~A.~Weaver\altaffilmark{22},
Pauline~Zarrouk\altaffilmark{25},
and Hu~Zou\altaffilmark{45}
}

\altaffiltext{1}{Lawrence Berkeley National Laboratory, 1 Cyclotron Road, Berkeley, CA 94720, USA}
\altaffiltext{2}{Institute for Astronomy, University of Edinburgh, Royal Observatory, Blackford Hill, Edinburgh EH9 3HJ, UK}
\altaffiltext{3}{Institute of Astronomy, University of Cambridge, Madingley Road, Cambridge CB3 0HA, UK}
\altaffiltext{4}{Space Telescope Science Institute, 3700 San Martin Drive, Baltimore, MD 21218, USA}
\altaffiltext{5}{Physics Dept., Boston University, 590 Commonwealth Avenue, Boston, MA 02215, USA}
\altaffiltext{6}{Dipartimento di Fisica ``Aldo Pontremoli'', Universit\`a degli Studi di Milano, Via Celoria 16, I-20133 Milano, Italy}
\altaffiltext{7}{Department of Physics \& Astronomy, University College London, Gower Street, London, WC1E 6BT, UK}
\altaffiltext{8}{Department of Physics and Astronomy, The University of Utah, 115 South 1400 East, Salt Lake City, UT 84112, USA}
\altaffiltext{9}{Instituto de F\'{\i}sica, Universidad Nacional Aut\'{o}noma de M\'{e}xico,  Cd. de M\'{e}xico  C.P. 04510,  M\'{e}xico}
\altaffiltext{10}{Department of Physics \& Astronomy and Pittsburgh Particle Physics, Astrophysics, and Cosmology Center (PITT PACC), University of Pittsburgh, 3941 O'Hara Street, Pittsburgh, PA 15260, USA}
\altaffiltext{11}{Center for Astrophysics $|$ Harvard \& Smithsonian, 60 Garden Street, Cambridge, MA 02138, USA}
\altaffiltext{12}{University of California, Berkeley, 110 Sproul Hall \#5800 Berkeley, CA 94720, USA}
\altaffiltext{13}{Institut de F\'{i}sica d’Altes Energies (IFAE), The Barcelona Institute of Science and Technology, Campus UAB, 08193 Bellaterra Barcelona, Spain}
\altaffiltext{14}{Departamento de F\'isica, Universidad de los Andes, Cra. 1 No. 18A-10, Edificio Ip, CP 111711, Bogot\'a, Colombia}
\altaffiltext{15}{Observatorio Astron\'omico, Universidad de los Andes, Cra. 1 No. 18A-10, Edificio H, CP 111711 Bogot\'a, Colombia}
\altaffiltext{16}{Institut d'Estudis Espacials de Catalunya (IEEC), 08034 Barcelona, Spain}
\altaffiltext{17}{Institute of Cosmology and Gravitation, University of Portsmouth, Dennis Sciama Building, Portsmouth, PO1 3FX, UK}
\altaffiltext{18}{Institute of Space Sciences, ICE-CSIC, Campus UAB, Carrer de Can Magrans s/n, 08913 Bellaterra, Barcelona, Spain}
\altaffiltext{19}{Fermi National Accelerator Laboratory, PO Box 500, Batavia, IL 60510, USA}
\altaffiltext{20}{Center for Cosmology and AstroParticle Physics, The Ohio State University, 191 West Woodruff Avenue, Columbus, OH 43210, USA}
\altaffiltext{21}{Department of Physics, The Ohio State University, 191 West Woodruff Avenue, Columbus, OH 43210, USA}
\altaffiltext{22}{NSF NOIRLab, 950 N. Cherry Ave., Tucson, AZ 85719, USA}
\altaffiltext{23}{Department of Physics, Southern Methodist University, 3215 Daniel Avenue, Dallas, TX 75275, USA}
\altaffiltext{24}{Department of Physics and Astronomy, University of California, Irvine, 92697, USA}
\altaffiltext{25}{Sorbonne Universit\'{e}, CNRS/IN2P3, Laboratoire de Physique Nucl\'{e}aire et de Hautes Energies (LPNHE), FR-75005 Paris, France}
\altaffiltext{26}{Department of Astronomy \& Astrophysics, University of Toronto, Toronto, ON M5S 3H4, Canada}
\altaffiltext{27}{Departament de F\'{i}sica, Serra H\'{u}nter, Universitat Aut\`{o}noma de Barcelona, 08193 Bellaterra (Barcelona), Spain}
\altaffiltext{28}{Department of Astronomy, The Ohio State University, 4055 McPherson Laboratory, 140 W 18th Avenue, Columbus, OH 43210, USA}
\altaffiltext{29}{Instituci\'{o} Catalana de Recerca i Estudis Avan\c{c}ats, Passeig de Llu\'{\i}s Companys, 23, 08010 Barcelona, Spain}
\altaffiltext{30}{Department of Physics and Astronomy, Siena College, 515 Loudon Road, Loudonville, NY 12211, USA}
\altaffiltext{31}{Department of Physics \& Astronomy, University  of Wyoming, 1000 E. University, Dept.~3905, Laramie, WY 82071, USA}
\altaffiltext{32}{Departamento de F\'{i}sica, Universidad de Guanajuato - DCI, C.P. 37150, Leon, Guanajuato, M\'{e}xico}
\altaffiltext{33}{Instituto Avanzado de Cosmolog\'{\i}a A.~C., San Marcos 11 - Atenas 202. Magdalena Contreras, 10720. Ciudad de M\'{e}xico, M\'{e}xico}
\altaffiltext{34}{IRFU, CEA, Universit\'{e} Paris-Saclay, F-91191 Gif-sur-Yvette, France}
\altaffiltext{35}{Department of Physics and Astronomy, University of Waterloo, 200 University Ave W, Waterloo, ON N2L 3G1, Canada}
\altaffiltext{36}{Perimeter Institute for Theoretical Physics, 31 Caroline St. North, Waterloo, ON N2L 2Y5, Canada}
\altaffiltext{37}{Waterloo Centre for Astrophysics, University of Waterloo, 200 University Ave W, Waterloo, ON N2L 3G1, Canada}
\altaffiltext{38}{Space Sciences Laboratory, University of California, Berkeley, 7 Gauss Way, Berkeley, CA  94720, USA}
\altaffiltext{39}{Instituto de Astrof\'{i}sica de Andaluc\'{i}a (CSIC), Glorieta de la Astronom\'{i}a, s/n, E-18008 Granada, Spain}
\altaffiltext{40}{Department of Physics and Astronomy, Sejong University, Seoul, 143-747, Korea}
\altaffiltext{41}{CIEMAT, Avenida Complutense 40, E-28040 Madrid, Spain}
\altaffiltext{42}{Department of Astrophysical Sciences, Princeton University, Princeton NJ 08544, USA}
\altaffiltext{43}{Department of Physics, University of Michigan, Ann Arbor, MI 48109, USA}
\altaffiltext{44}{University of Michigan, Ann Arbor, MI 48109, USA}
\altaffiltext{45}{National Astronomical Observatories, Chinese Academy of Sciences, A20 Datun Rd., Chaoyang District, Beijing, 100012, P.R. China}

\thanks{$^*$E-mail: \href{mailto:rongpuzhou@lbl.gov}{rongpuzhou@lbl.gov}}

\begin{abstract}
We present new Galactic dust reddening maps of the high Galactic latitude sky using DESI imaging and spectroscopy. We directly measure the reddening of 2.6 million stars by comparing the observed stellar colors in $g-r$ and $r-z$ from DESI imaging with the synthetic colors derived from DESI spectra from the first two years of the survey. The reddening in the two colors is on average consistent with the \cite{fitzpatrick_correcting_1999} extinction curve with $R_\mathrm{V}=3.1$. We find that our reddening maps differ significantly from the commonly used \cite{schlegel_maps_1998} (SFD) reddening map (by up to 80 mmag in $E(B-V)$), and we attribute most of this difference to systematic errors in the SFD map. To validate the reddening map, we select a galaxy sample with extinction correction based on our reddening map, and this yields significantly better uniformity than the SFD extinction correction. Finally, we discuss the potential systematic errors in the DESI reddening measurements, including the photometric calibration errors that are the limiting factor on our accuracy. The $E(g-r)$ and $E(r-z)$ maps presented in this work, and for convenience their corresponding $E(B-V)$ maps with SFD calibration, are publicly available.

\end{abstract}


\section{Introduction} \label{sec:intro}

\setcounter{footnote}{0}  

Galactic dust absorbs and scatters light mostly in UV and optical wavelengths and re-emits in the infrared as thermal radiation. The dust extinction in the optical wavelengths can be a significant nuisance for galaxy surveys as it makes sources behind the dust appear dimmer and redder. The intrinsic brightness and color of an extragalactic source are usually inferred by applying a correction to the observed brightness and color. The amplitude of this correction is often calculated using a reddening map that provides the reddening for a specific wavelength or color and an extinction curve (or ``extinction law'') that prescribes the amount of extinction as a function of wavelength.

Accurate extinction correction, along with accurate photometric calibration, is essential for precision cosmology (e.g., see \citealt{ross_clustering_2012,huterer_calibration_2013,thelsstdarkenergysciencecollaboration_lsst_2021}). It ensures the angular uniformity of the inferred intrinsic photometry, which in turn ensures the uniformity of the galaxy samples selected with the extinction-corrected photometry. Extinction correction also affects the absolute photometric calibration, which is crucial for standard candles such as Type Ia supernovae.

The reddening map from \cite{schlegel_maps_1998} (hereafter SFD) was an indirect measurement inferred from the thermal dust emission in the infrared, and it has been widely used in galaxy surveys.
However, the infrared emission-based reddening maps rely on the conversion from thermal radiation to optical absorption that may be inaccurate, and they are also susceptible to contamination from the cosmic infrared background (e.g., see \citealt{chiang_extragalactic_2019,chiang_corrected_2023}). \cite{lenz_new_2017} presented an alternative reddening map based on HI 21 cm emission, but it is also an indirect measurement subject to modeling errors and is limited to low HI column densities.

Direct reddening measurements of individual objects (stars or galaxies) are also possible, provided that their intrinsic spectra or colors are known, and this direct approach has the advantage that it is not affected by the astrophysical uncertainties in the indirect measurements. Earlier work such as \cite{schlafly_measuring_2011} based on stars and \cite{peek_correction_2010} based on galaxies demonstrated the feasibility of creating reddening maps via this approach. Recent advances in wide-field imaging and spectroscopic surveys are now enabling the direct measurements of Galactic reddening with precision and resolution comparable or even superior to emission-based maps such as SFD.

Here we present a new Galactic reddening map based on direct stellar reddening measurements using imaging and spectroscopy of the Dark Energy Spectroscopic Instrument (DESI) \citep{DESI2016a.Science,DESI2016b.Instr}. For this measurement, we take advantage of more than two million stellar spectra obtained during the first two years of the DESI survey. We use the equivalent widths and profiles of absorption lines present in the spectra of stars to fit a stellar atmosphere model for each of them, independently from any knowledge about their spectro-photometric calibration. We then use those models to compute synthetic colors and compare them with measurements from imaging surveys. The difference between the observed and synthetic colors provides us with information about reddening by Galactic dust.

In this paper, we measure the reddening, i.e., the change in the colors caused by Galactic extinction. While the amount of extinction, i.e., the change in flux or magnitude, can be inferred from the reddening value by assuming some extinction curve, we do not attempt to directly measure extinction which would require extra information such as accurate distances of the stars.

And while we are able to measure the variation of the reddening across the sky with good accuracy, the absolute zero point in the reddening measurement is somewhat uncertain as it depends on the accuracies of the stellar models and the photometric calibration, both of which have significant uncertainties. In other words, the measured reddening can be different from the truth by an unknown additive constant. Rather than attempting to determine reddening zero point, we adopt the SFD zero point in this work (in the procedure described in Section \ref{sec:calibration}). It's worth noting that for galaxy sample selection, the absolute zero point level does not affect uniformity. One potential future improvement is to use specific categories of stars for this purpose, such as pure hydrogen white dwarfs (e.g., in \citealt{manser_desi_2024}), for which we have possibly more accurate synthetic spectra.

The paper is structured as follows. Section \ref{sec:data} describes the imaging and spectroscopic data and the selection of the stellar sample. Section \ref{sec:measurements} describes how we measure the reddening (\S\ref{sec:method}), calibrate the systematic offsets in the synthetic colors (\S\ref{sec:calibration}), estimate the measurement errors (\S\ref{sec:error_estimation}), and produce the reddening map using individual stellar measurements (\S\ref{sec:reddening_maps}).
Section \ref{sec:discussion} discusses the results: we compare the DESI map with SFD which reveals significant differences (\S\ref{sec:sfd_comparison}); we validate the two maps by checking the uniformity of a galaxy sample selected with extinction-corrected photometry based on the two maps (\S\ref{sec:elg_selection}); we validate the extinction curve by comparing the reddening in two colors (\S\ref{sec:color-color}). Section \ref{sec:systematics} discusses the potential sources of systematic errors. We describe the data products in Section \ref{sec:data_availability}, and we conclude in Section \ref{sec:conclusion}.

\section{Data} \label{sec:data}

We use the Legacy Surveys (LS) DR9 imaging\footnote{\url{https://www.legacysurvey.org/dr9/}} \citep{dey_overview_2019,schlegel_ls_dr9} for the observed stellar photometry in $g$, $r$ and $z$ bands. The LS imaging consists of two regions, namely North and South, that are observed with different telescopes and are separated at declination of approximately $32^{\circ}$. The Southern imaging is provided by the Dark Energy Camera Legacy Survey (DECaLS; \citealt{dey_overview_2019}) from the Cerro Tololo Inter-American Observatory using the DECam instrument at the Blanco telescope, with additional DECam data from other programs including the Dark Energy Survey \citep{darkenergysurveycollaboration_dark_2016}. The Northern imaging is provided by two surveys at the Kitt Peak National Observatory: the Beijing–Arizona Sky Survey (BASS; \citealt{zou_project_2017}) provides $g$ and $r$ band imaging from the Bok telescope, and the Mayall $z$-band Legacy Survey (MzLS; \citealt{dey_overview_2019}) provides $z$-band imaging from the Mayall telescope.

We use the DESI spectroscopic data from the first two years of observations, from May 14, 2021, to June 21, 2023, to obtain the synthetic stellar spectra and the resulting synthetic colors. We use the Y1 reduction (internally named the ``iron'' data release) for the first-year data (taken from May 14, 2021, to June 14, 2022) and the internal ``daily'' reduction of the second-year data (taken from the end of Y1 to June 21, 2023).  Specifically, for the first-year data we use the HEALPix \footnote{\url{https://healpix.sourceforge.io/}} coadds (which co-add across overlapping tiles; see \S3.3.2 of \citealt{DESI_EDR}), and for the second-year data we use the tile-based coadds (in which repeat observations in overlapping tiles are not combined).

The stellar sample includes stars observed in the dark, bright and backup programs of the DESI Main Survey (see \citealt{schlafly_survey_2023}). We select the sample using the \texttt{desitarget} bitmask (see \citealt{myers_targetselection_2023}):
\begin{enumerate}
\item Standard stars in all three programs (STD\_FAINT and STD\_BRIGHT in dark and bright programs; GAIA\_STD\_FAINT and GAIA\_STD\_BRIGHT in the backup program); see also \cite{guy_spectroscopic_2023} and \cite{myers_targetselection_2023};
\item Milky Way Survey (MWS) magnitude limited blue sample, red sample, and bulk sample in the bright program (MWS\_MAIN\_BLUE, MWS\_MAIN\_RED, and MWS\_BROAD); see also \cite{cooperOverviewDESIMilky2023};
\item Brighter and fainter backup targets in the backup program (BACKUP\_BRIGHT and BACKUP\_FAINT); see also \cite{deyBackupProgramDark2025}.
\end{enumerate}

The DESI stars are cross-matched to both LS DR9 and Gaia DR3 with a 0.1 arcsec search radius (with correction for proper motion). While most of the stars are selected with LS DR9, the stars in the DESI backup program are selected with Gaia and are not associated with LS imaging, thus necessitating the cross-matching to the LS. Cross-matching to LS DR9 also allows us to measure the reddening in the Southern filters for stars targeted with Northern photometry and vice versa in the overlapping region; we use the overlapping region to measure the reddening offsets between the two regions. Moreover, LS DR9 only contains Gaia DR2 data, and the cross-matching allows us to use the more recent Gaia DR3 data. The faintest star in our sample has Gaia $G \approx 19.5$, which is much brighter than the Gaia limiting magnitude of $G \approx 21$.

We apply a variety of selection cuts to obtain our final sample; see Section \ref{sec:quality_cuts} for details. Furthermore, we remove duplicates (caused by either repeated observations of a source in different survey/program or because of the observations in overlapping tiles in Y2 data) by keeping objects with the highest signal-to-noise ratio (S/N), specifically the median per-angstrom S/N in the blue spectrograph (hereafter referred to as SN\_B).

\begin{figure*}
    \centering
    \includegraphics[width=2.1\columnwidth]{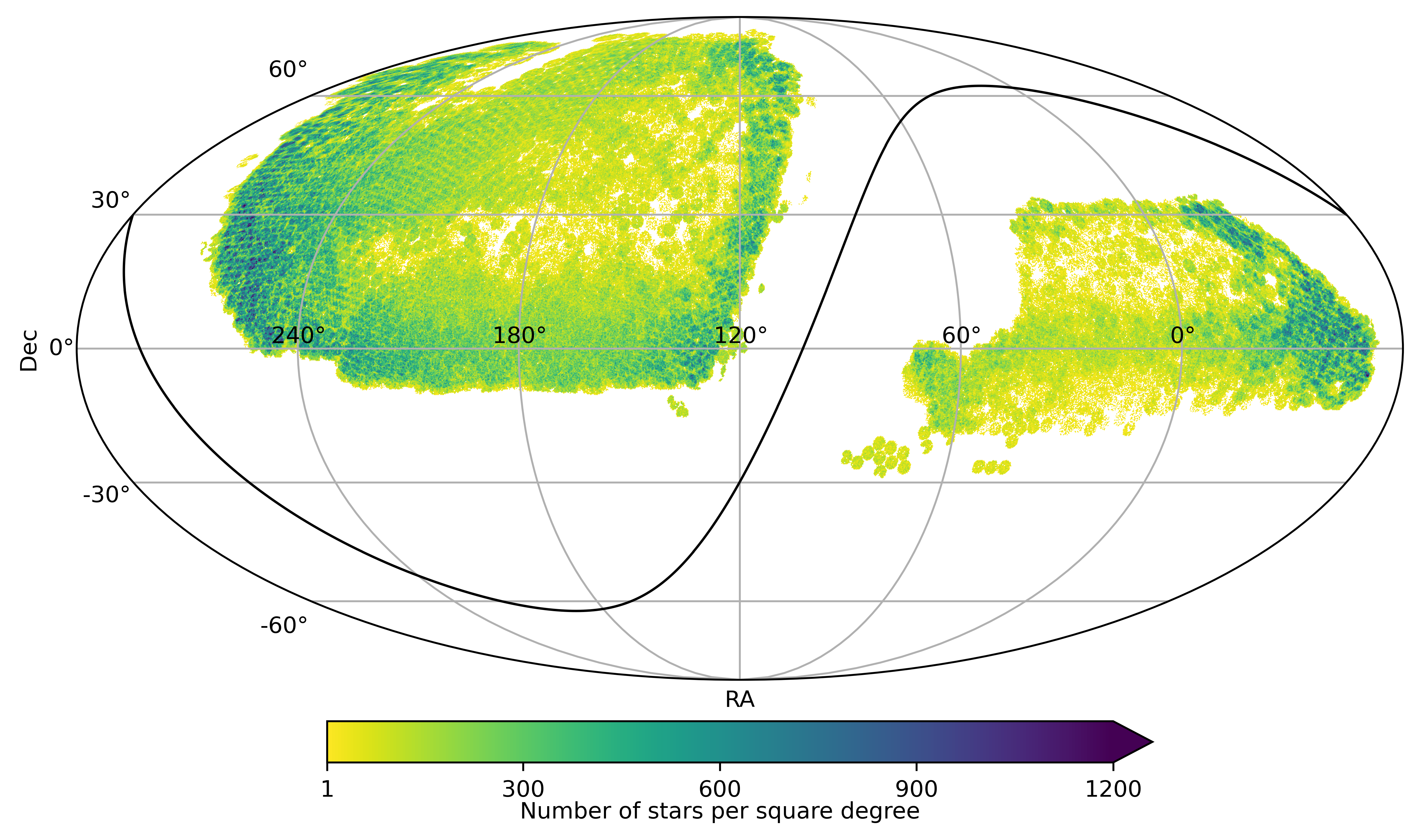}
    \caption{The density of stars used in $g-r$ reddening measurements (see Section \ref{sec:measurements}). The gaps and missing pixels are due to the lack of coverage from the first two years of DESI observations. The curve separating the two regions is the Galactic plane. The HEALPix resolution is NSIDE=256 (13.7$'$ pixel size). The color scale is linear in this figure and all other HEALPix map figures.}
    \label{fig:n_star_map}
\end{figure*}

Table \ref{tab:summary_stats} lists the number of stars from each target type in the final sample. Figure \ref{fig:n_star_map} shows the density map of the final stellar sample. The total sky coverage is 13,835 square degrees, based on the number of HEALPix pixels with NSIDE=256 (13.7$'$ pixel size) that contain at least one star in the $g-r$ reddening map.

\begin{table*}
    \centering
    \begin{tabular}{llllll}
    \hline
    Program & Target class & $N_{g-r}$ & $N_{r-z}$ & $\sigma(E(g-r))$ & $\sigma(E(r-z))$\\
    \hline
    Dark & STD\_BRIGHT & 134695 & 128377 & 15 (13) & 14 (13) \\
    Dark & STD\_FAINT & 269140 & 261834 & 15 (13) & 15 (13) \\
    Dark & All & 269140 & 261834 & 15 (13) & 15 (13) \\
    \hline
    Bright & STD\_BRIGHT & 117124 & 108658 & 16 (14) & 15 (13) \\
    Bright & STD\_FAINT & 239596 & 229411 & 18 (16) & 17 (15) \\
    Bright & MWS\_MAIN\_BLUE & 1922429 & 1760952 & 20 (16) & 22 (18) \\
    Bright & MWS\_MAIN\_RED & 62498 & 56549 & 27 (19) & 30 (22) \\
    Bright & MWS\_BROAD & 35074 & 25289 & 25 (15) & 28 (19) \\
    Bright & All & 2021800 & 1844168 & 20 (16) & 22 (19) \\
    \hline
    Backup & GAIA\_STD\_BRIGHT & 133834 & 118975 & 18 (15) & 19 (16) \\
    Backup & GAIA\_STD\_FAINT & 143351 & 128526 & 19 (15) & 19 (17) \\
    Backup & BACKUP\_BRIGHT & 155861 & 45164 & 17 (13) & 18 (15) \\
    Backup & BACKUP\_FAINT & 77123 & 54310 & 17 (13) & 20 (18) \\
    Backup & All & 345433 & 205900 & 18 (14) & 19 (17) \\
    \hline
    All & All & 2636373 & 2311902 & 20 (15) & 22 (18) \\
    \hline
    \end{tabular}
    \caption{Statistics of the different types of stars used for the reddening measurement, i.e., after the selection cuts (see text for details). The number of stars ($N_{g-r}$ and $N_{r-z}$) is the total number of stars used in the $g-r$ and $r-z$ reddening measurements (after passing the quality cuts). There are 2205615 stars that pass both the $g-r$ and $r-z$ quality cuts. Note that there is overlap between some of the target classes (e.g., STD\_BRIGHT is entirely a subset of STD\_FAINT). The $\sigma(E(g-r))$ and $\sigma(E(r-z))$ columns list the per-star error in mmag (millimagnitude) in the reddening measurements; the errors are obtained by comparing the per-star measurements with the map pixel values (which average over the reddening measurement of at least 32 stars) of NSIDE=256 (13.7$'$ pixel size) maps; the first number is the root-mean-square (RMS) error, and the second number in parentheses is normalized median absolute deviation (NMAD), here defined as $\sigma_\mathrm{NMAD}(\Delta x)=1.4826 \times \mathrm{Median}(|\Delta x|)$. Some target classes have much larger RMS errors than NMAD errors because they have a small fraction of stars with large errors (to which $\sigma_\mathrm{NMAD}$ is not sensitive).}
    \label{tab:summary_stats}
\end{table*}

\subsection{Modeling the stellar spectra: rvspecfit}
\label{sec:rvspecfit}

We model the observed spectra using the template fitting code \texttt{rvspecfit} \citep{koposov_rvspecfit_2019, koposov_accurate_2011} to obtain the synthetic colors. 
\texttt{rvspecfit} was designed to fit non-flux calibrated spectra. Thus it models the observed spectra $S(\lambda)$ by $T(\lambda \sqrt{\frac{1+v/c}{1-v/c}},p)*P(\lambda)$ where $T(\lambda, p)$ is the interpolated stellar spectrum for a given set of stellar parameters $p$, $v$ is the radial velocity, and $P(\lambda)$ is either a polynomial or linear combination of any set of basis functions (such as radial basis functions (RBFs)). For DESI, $P(\lambda)$ currently consists of a 2nd order polynomial (1, $\lambda$, $\lambda^2$) and 7 Gaussian RBFs centered at equally spaced points along wavelength. The standard deviation of each Gaussian RBF is 1/7 of the wavelength range. 

Since $P(\lambda)$ is a linear combination of fixed functions, the best values of the linear coefficients (in a $\chi^2$ sense) for a given $P(\lambda)$ can be found through a simple matrix operation (see \citealt{koposov_accurate_2011} for more details). Thus, when fitting the stellar spectra, the optimization is performed over $P(\lambda)$ and radial velocity using a combination of Nelder-Mead and BFGS optimizers (see also \citealt{cooperOverviewDESIMilky2023}). (For simplicity, we refer to $P(\lambda)$ as ``the polynomial term''.)

The grid of stellar spectra used by \texttt{rvspecfit} is the PHOENIX v2.0 grid by \cite{husser_new_2013} convolved to the average DESI resolution of $\mathrm{FWHM}=1.55\textup{~\AA}$ for blue and red cameras and $\mathrm{FWHM}=1.8\textup{~\AA}$ for the near-infrared camera.

This work relies on the same \texttt{rvspecfit} version that was used for the processing of the Milky Way Survey data for Early Data Release catalog \citep{desi_edr_mws_vac}. That version suffers from gridding effects where the stellar parameters tend to cluster around nodes of the original grid of templates (see Figure 9 in \citealt{desi_edr_mws_vac} and our Figure \ref{fig:calibration}). The first DESI Data Release will rely on the processing by the newer version of \texttt{rvspecfit} with an improved interpolation scheme that solves the gridding problem.

\begin{figure*}
    \centering
    \includegraphics[width=1.9\columnwidth]{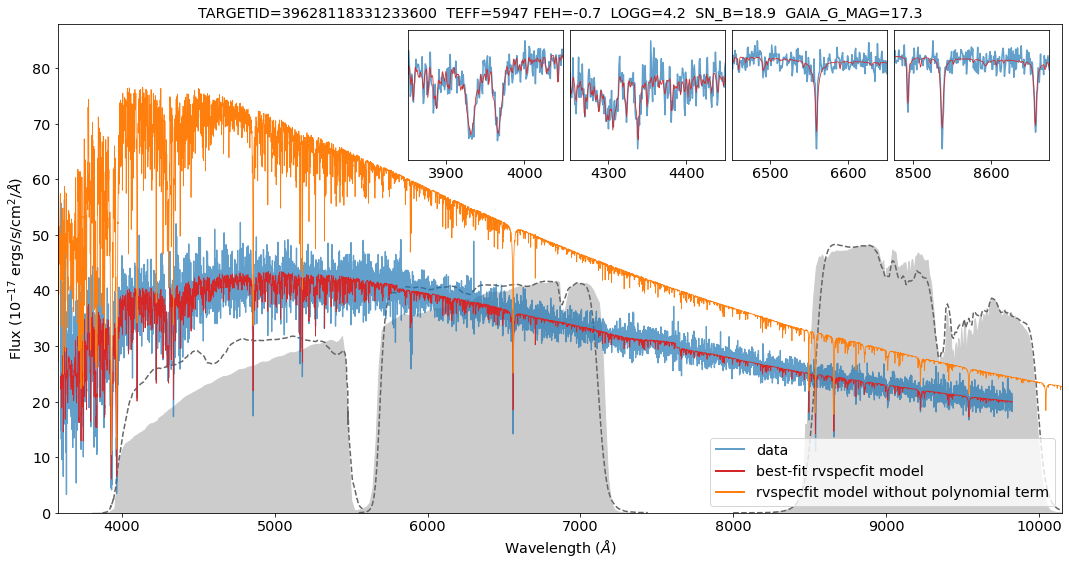}
    \caption{An example DESI spectrum: a MWS\_MAIN\_BLUE target with $E(g-r)=0.192$ ($E(B-V)_\mathrm{SFD}=0.162$). The spectrum has a per-angstrom S/N of 18.9 in the blue spectrograph, compared to a median S/N of 18.2 for the full sample. The observed spectrum is in blue and the best-fit \texttt{rvspecfit} model is in red. The insets zoom in on some of the absorption lines. The synthetic spectrum without the polynomial term $P(\lambda)$ (see Section \ref{sec:rvspecfit}), shown in orange, is the zero-extinction spectrum and is used to compute the intrinsic color. The synthetic spectra extend redder than the observed spectra and cover the entire wavelength range of the $z$ filter. For illustrative purposes, the zero-extinction spectrum is normalized such that its total flux matches that of the extinction-corrected observed spectrum. The $g$, $r$ and $z$ filter curves, used for synthesizing the photometry, are shown as the gray-filled curves for DECam filters (Southern imaging) and dashed curves for BASS and MzLS filters (Northern imaging) with arbitrary normalization.}
    \label{fig:example_spectrum}
\end{figure*}

For each star \texttt{rvspecfit} produces a best-fit synthetic spectrum with zero extinction as well as the following stellar atmosphere parameters that determine the synthetic spectrum: effective temperature $T_\mathrm{eff}$, surface gravity log $g$, metallicity [Fe/H], and alpha element abundance [$\alpha$/Fe]. Figure \ref{fig:example_spectrum} shows an example DESI spectrum and the \texttt{rvspecfit} best-fit model. We use the best-fit spectral models without the polynomial term to compute zero-extinction synthetic colors and compare them with the observed colors from imaging to determine the Galactic reddening. Given this objective, it is important to note that because of the multiplicative polynomial term $P(\lambda)$, the stellar parameters mostly only depend on the narrow absorption features, and the actual amount of extinction has little impact on the best-fit model (at least for the typical amount of extinction, $E(B-V)\lesssim 0.2$, encountered by most cosmological probes). We will discuss this more in Section \ref{sec:model_extinction_bias}. This polynomial term removes the dependence of not only extinction but also other systematic errors such as flux calibration errors in DESI spectra.

For each star, we obtain zero-extinction synthetic colors by integrating the synthetic spectra (without the multiplicative polynomial) with the filter curves. The Southern and Northern imaging are based on different instruments, and the appropriate filter curves for each imaging region are used. We also compute the SFD-reddened synthetic magnitudes by artificially applying the extinction on the synthetic spectra with SFD $E(B-V)$ value; the SFD-reddened colors are only used in the initial step of calibrating the synthetic colors (see Section \ref{sec:calibration}).

\subsection{Quality cuts}
\label{sec:quality_cuts}

We apply quality cuts to select stars that yield reliable reddening measurements. The full list of the quality cuts is in Table \ref{tab:quality cuts}. We describe the quality cuts below.

\begin{table*}
\label{tab:quality_cuts}
    \centering
    \label{tab:quality cuts}
    \begin{tabular}{ll}
    \hline
    Cuts & Comment \\
    \hline
    
    \hline
    \multicolumn{2}{c}{Parallax and imaging quality cuts}\\
    PARALLAX $<$ 1 \& PARALLAX\_ERROR$\neq$0 & Require parallax less than 1 mas and valid parallax value (see caption) \\
    NOBS\_G$>$0 \& NOBS\_R$>$0 \& NOBS\_Z$>$0 & Require at least 1 observation in $g$, $r$ and $z$ bands\\
    FLUX\_IVAR\_G$>$0 \& FLUX\_IVAR\_R$>$0 & \multirow{2}{25em}{Require valid fluxes in $g$, $r$ and $z$ bands}\\
    \ \& FLUX\_IVAR\_Z$>$0 \\
    PHOT\_G\_MEAN\_MAG$\neq$0 \& PHOT\_BP\_MEAN\_MAG$\neq$0 & \multirow{2}{25em}{Require valid Gaia magnitudes in G, BP and RP bands}\\
    \ \& PHOT\_RP\_MEAN\_MAG$\neq$0 \\
    
    \hline
    \multicolumn{2}{c}{Quality cuts on the $g-r$ sample}\\
    ANYMASK\_G=0 \& ANYMASK\_R=0 & Remove stars flagged by imaging pipeline (e.g., due to saturation)\\
    FRACFLUX\_G$<$0.01 \& FRACFLUX\_R$<$0.01 & Remove blended stars based on FRACFLUX\\
    FRACMASKED\_G$<$0.6 \& FRACMASKED\_R$<$0.6 & Remove stars with a large fraction of masked pixels\\
    FIBERFLUX\_G/FIBERTOTFLUX\_G$>$0.99 & \multirow{2}{25em}{Remove blended stars based on fiber flux fraction}\\
    \ \& FIBERFLUX\_R/FIBERTOTFLUX\_R$>$0.99 & \\
    $|g_\mathrm{LS} - g_\mathrm{Gaia}| < 0.1$  \& $|r_\mathrm{LS} - r_\mathrm{Gaia}| < 0.1$ & Remove outliers by comparing with Gaia photometry\\
        \multirow{2}{25em}{$\sigma_{g-r,\mathrm{NN}} < 0.03$} & Remove stars with large reddening errors based on 400 nearest neighbors \\
    & in stellar parameter space\\
    
    \hline
    \multicolumn{2}{c}{Quality cuts on the $r-z$ sample}\\
    ANYMASK\_R=0 \& ANYMASK\_Z=0 & Remove stars flagged by imaging pipeline (e.g., due to saturation)\\
    FRACFLUX\_R$<$0.01 \& FRACFLUX\_Z$<$0.01 & Remove blended stars based on FRACFLUX\\
    FRACMASKED\_R$<$0.6 \& FRACMASKED\_Z$<$0.6 & Remove stars with a large fraction of masked pixels\\
    FIBERFLUX\_R/FIBERTOTFLUX\_R$>$0.99 & \multirow{2}{25em}{Remove blended stars based on fiber flux fraction}\\
    \ \& FIBERFLUX\_Z/FIBERTOTFLUX\_Z$>$0.99 & \\
    $|r_\mathrm{LS} - r_\mathrm{Gaia}| < 0.1$  \& $|z_\mathrm{LS} - z_\mathrm{Gaia}| < 0.1$ & Remove outliers by comparing with Gaia photometry\\
    \multirow{2}{25em}{$\sigma_{r-z,\mathrm{NN}} < 0.035$} & Remove stars with large reddening errors based on 400 nearest neighbors \\
    & in stellar parameter space\\

    \hline
    \multicolumn{2}{c}{Spectroscopic quality cuts}\\
    EFFTIME\_SPEC $>$ 40 (applied on tiles) & Remove tiles with low effective exposure time \\
    EFFTIME\_LRG$>$30 (applied on objects) & Remove spectra with low effective exposure time \\
    SN\_B $>$ 5 & Minimum median per-angstrom S/N in the blue spectrograph arm\\
    ZWARN=0 & Remove problematic spectra flagged by the ZWARNING bitmask \\
    DELTACHI2$>$100 & Remove spectra with low $\Delta \chi^2$ value from \texttt{redrock}\\
    SPECTYPE=STAR & Remove galaxies or quasars\\
    SUBTYPE $\neq$ WD & Remove objects classified as white dwarfs by \texttt{redrock}\\
    RVS\_WARN=0 & Remove problematic spectra marked by the \texttt{rvspecfit} warning flag \\
    Z$<$0.002 & Additional cut to remove galaxies \\

    \hline
    \multicolumn{2}{c}{Stellar atmosphere parameter cuts}\\
    3.5 $<$ LOGG $<$ 5.2 & Surface gravity log $g$ \\
    5000 $<$ TEFF $<$ 6500 & Effective temperature $T_\mathrm{eff}$ \\
    -4.0 $<$ FEH $<$ 0.5 & Metallicity [Fe/H] \\
    \hline
    \end{tabular}
    \caption{List of quality cuts that we apply to select stars with reliable reddening measurements. The quantity names in all caps are column names in the catalog. $g_\mathrm{LS}$, $r_\mathrm{LS}$, $z_\mathrm{LS}$ are the observed LS magnitudes; $g_\mathrm{Gaia}$, $r_\mathrm{Gaia}$, $z_\mathrm{Gaia}$ are Gaia magnitudes (G, BP and RP) transformed into LS filters. The imaging and spectroscopic cuts ensure the quality of the input data, and the stellar parameter cuts select the region of parameter space where the stellar atmosphere model yields reliable broad-band colors. The Gaia catalogs ingested by the Legacy Surveys have zero-filled values for invalid measurements; thus we require, e.g., PARALLAX\_ERROR$\neq$0 to remove objects with invalid measurements.}
\end{table*}

We first remove stars within 1 kpc from us by requiring that the Gaia parallax be less than 1 mas (milliarcsecond). The DESI footprint is almost entirely above the galactic latitude of $|b|=18^{\circ}$. This corresponds to a distance of 310 pc from the Galactic plane at the lowest latitude in the survey, which is significantly larger than the typical dust scale height of $\sim$100 pc, ensuring that the stars used are behind most of the dust. While the 1 kpc limit should be adequate for most of the high Galactic latitude sky, we will investigate more optimized parallax cuts in future studies.

For the imaging, we require at least one valid observation in the $g$, $r$ and $z$ bands, and we also require valid Gaia magnitudes in G, BP and RP bands. We then use the imaging quality values and flags to remove any stars that may have unreliable colors or spectra, e.g., due to saturation or blending. Different quality cuts are applied for $g-r$ color and $r-z$ color. For instance, a star with bad $z$-band imaging is retained for the $g-r$ sample if it meets the $g-r$ quality cuts. As a result, the $g-r$ and $r-z$ samples have slightly different sets of stars, and we treat them as two separate samples. The difference is largest in the Northern imaging where $z$-band saturates more easily, resulting in lower density in the $r-z$ sample in that region. The imaging quality cuts remove 21\% of the $g-r$ sample in the North and 14\% in the South; for the $r-z$ sample, 42\% in the North and 20\% are removed. Most of these stars are removed due to saturation.

We also try to identify and remove photometric outliers by comparing with Gaia photometry: we use Gaia magnitudes in G, BP and RP bands to predict magnitudes in LS filters, and we remove any stars whose observed magnitude differs from the predicted magnitude by more than 0.1 mag. This removes less than 1\% of the stars. Details of the Gaia-to-LS transformation are described in Appendix \ref{sec:gaia-to-ls}. Roughly 40\% of the stars removed by this cut have Gaia Renormalized Unit Weight Error (RUWE)\footnote{Document GAIA-C3-TN-LU-LL-124-01 from \url{https://www.cosmos.esa.int/web/gaia/public-dpac-documents}} larger than 1.4. For comparison, less than 1\% of the remaining stars have RUWE$>$1.4. Gaia's recommended criterion of good astrometric solutions of single stars is RUWE$<$1.4. Therefore, many (if not most) of the stars removed by the Gaia photometric outlier cut may be double stars.

We also apply spectroscopic quality cuts to remove spectra with low S/N or flagged by the DESI pipeline \citep{guy_spectroscopic_2023}, specifically the redshift fitter \texttt{redrock}\footnote{\url{https://github.com/desihub/redrock}} \citep{bailey_redrock_2024}. We require that the object is classified as a star, and that it is not a white dwarf (which are not included in the PHOENIX synthetic spectra and thus cannot be accurately modeled by \texttt{rvspecfit}).

Finally, we select stars within a certain range of stellar atmosphere parameters. The limits on the stellar parameters are decided empirically based on the accuracy of the synthetic colors from the \texttt{rvspecfit} model. Stars with parameters outside the range typically have significantly larger errors in their synthetic color than those within the range; see the upper panel of Figure \ref{fig:color_offset_trends_before}. The stellar parameter cuts remove roughly 23\% of the stars (most of the removed stars are below the $T_\mathrm{eff}$ threshold of 5000 K). The selected temperature range of 5000--6500 K roughly corresponds to G and F spectral types. While the stars removed by the stellar parameter cuts have larger errors and/or systematic offsets, some of them may still be useful for reddening measurements, and we will explore the broader stellar parameter space in future studies.

Overall, the quality cuts remove about 16\% of the stars observed in the dark program, 50\% in the bright program, and 70\% in the backup program. The rejection rates of the bright and backup programs are much higher than that of the dark program because 1) the stars observed in the dark program are standard stars which are selected to be F/G type stars, whereas most of the stars in the bright and backup programs have a much broader selection, and a significant fraction of them are below the $T_\mathrm{eff}$ threshold; 2) the bright and backup stars are brighter and thus a larger fraction of them are saturated in the imaging.

\section{Reddening measurements}
\label{sec:measurements}

In this section, we describe how we measure the reddening using the aforementioned dataset and create the reddening maps.

\subsection{Method}
\label{sec:method}

We directly measure the reddening in the color $a-b$ (either $g-r$ or $r-z$ in our dataset)
\begin{equation}
\label{eq:delta_color}
E(a-b) = (a-b)_\mathrm{obs} - (a-b)_{E=0}
\end{equation}
where $(a-b)_\mathrm{obs}$ is the observed color and $(a-b)_{E=0}$ is the synthetic color with no extinction.

If we assume an extinction curve, the amount of extinction at some wavelength for a given line of sight can be expressed as the wavelength-dependent extinction coefficient $R(\lambda)$ multiplied by the reddening $E$ in a pre-defined color at that line of sight
\begin{equation}
\label{eqn:extinction}
A(\lambda) \equiv m_{\lambda,\mathrm{obs}} - m_{\lambda,E=0} \equiv R(\lambda) E
\end{equation}
where $m_{\lambda,\mathrm{obs}}$ is the observed magnitude at that wavelength, and $m_{\lambda,E=0}$ is the magnitude if there is no extinction. $R(\lambda)$ is the extinction coefficient and is specified by the extinction curve. Similarly, we can write the extinction for a filter  $x$ with a finite bandpass as\footnote{Note that the second equality is an approximation for broad-band filters as $R_x$ will be different at high extinction.}
\begin{equation}
\label{eqn:rx}
A_x \equiv m_{x,\mathrm{obs}} - m_{x,E=0} \equiv R_x E
\end{equation}

The definition of $E$ is arbitrary. For historical reasons, the reddening maps often report the reddening $E$ in the $B-V$ color. Here we adopt this definition of $E$ when reporting the $R_x$ values. Specifically, we adopt the calibration of the SFD $E(B-V)$ map\footnote{As \cite{schlafly_measuring_2011} has pointed out, a recalibration is needed to bring the SFD $E(B-V)$ to the $B-V$ reddening in actual $B$ and $V$ filters. But since the exact definition of $E$ has no effect on our $E(g-r)$ and $E(r-z)$ measurements, we simply adopt the original SFD calibration.}. The conversion between $E(a-b)$ and $E(B-V)$ is
\begin{equation}
E(a-b) = \left( R_a - R_b \right) E(B-V)\\ \label{eq:ebv}
\end{equation}

\begin{table}
    \centering
    \begin{tabular}{lll}
    \hline
    Filter & $R_x$ & $\lambda_\mathrm{eff}$ (\AA)\\
    \hline
    DECam $g$ & 3.214 & 4827.1 \\
    DECam $r$ & 2.165 & 6382.7 \\
    DECam $z$ & 1.211 & 9105.8 \\
    BASS $g$ & 3.258 & 4777.3 \\
    BASS $r$ & 2.176 & 6360.9 \\
    MzLS $z$ & 1.199 & 9158.4 \\
    \hline
    \end{tabular}
    \caption{Extinction coefficients and effective wavelengths for the Southern imaging (DECam filters) and Northern imaging (BASS and MzLS filters). The DECam coefficients are the same as those adopted by LS DR9.}
    \label{tab:filters}
\end{table}

To convert our reddening measurements to $E(B-V)$, we assume a universal extinction curve of \cite{fitzpatrick_correcting_1999} with $R_\mathrm{V}=3.1$. The extinction coefficients $R_x$ are computed based on an F dwarf star with 7000 K and [Fe/H]=$-1$. Table \ref{tab:filters} lists the extinction coefficients for our filters. The table also lists the effective wavelengths, which are defined as the wavelength where the extinction coefficient equals that of the filter, i.e., $R(\lambda_\mathrm{eff})=R_x$. (This definition is allowed because the extinction curve is monotonic in the optical wavelengths and thus invertible.) The extinction coefficients are slightly different for the Northern and Southern imaging due to differences in the filter curves and (to a lesser extent) in the average airmass. For the DECam filters (Southern imaging), we assume an airmass of 1.3. For the BASS and MzLS filters (Northern imaging), we assume an airmass of 1.1.

Since it is inconvenient to keep track of two slightly different filter sets for the reddening maps, we renormalize the Northern reddening measurements to match the $R_a - R_b$ value of the South before producing the reddening maps and only report the reddening as would be measured the DECam filters (see Section \ref{sec:reddening_maps}).

We infer $E(B-V)$ (with SFD normalization) from $E(g-r)$ and $E(r-z)$ (in DECam filters) with the following conversions:
\begin{align}
E(B-V) = E(g-r) / 1.049 \label{eq:egr_to_ebv}\\
E(B-V) = E(r-z) / 0.954 \label{eq:erz_to_ebv}
\end{align}
where 1.049 and 0.954 are $R_g - R_r$ and $R_r - R_z$, respectively.

Note that while we assume a universal extinction curve to infer $E(B-V)$, our $E(g-r)$ and $E(r-z)$ measurements are direct measurements and they do not rely on assumptions of the extinction curve (except for a slight dependence in the renormalization the Northern reddening measurements).

\subsection{Calibration of synthetic colors}
\label{sec:calibration}

\begin{figure*}
    \centering
    \includegraphics[width=2.1\columnwidth]{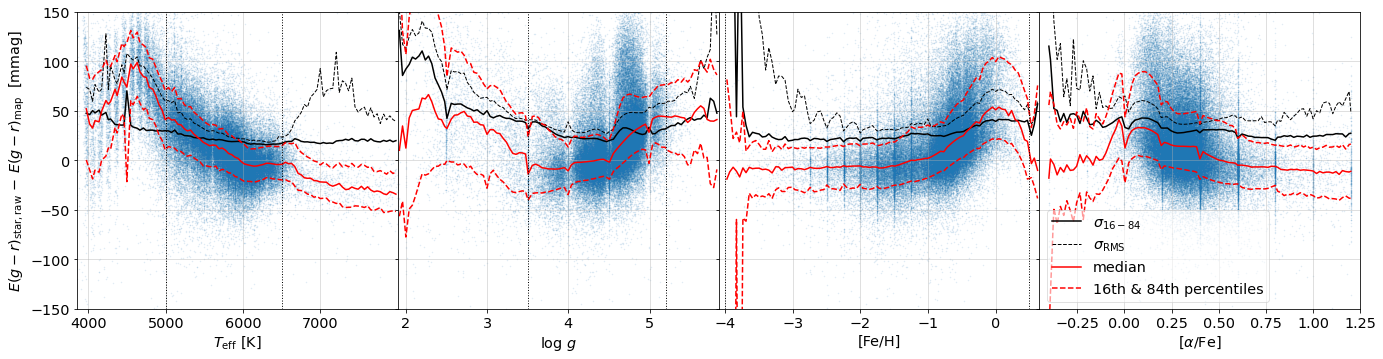}
    \includegraphics[width=2.1\columnwidth]{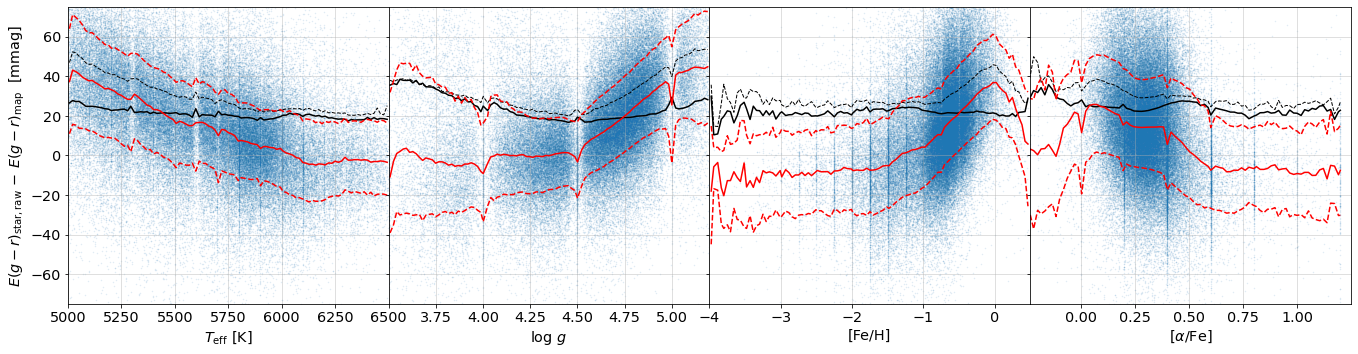}
    \caption{Color offset trends in $g-r$ before calibration for bright-time MWS stars in the South. Upper panel: stars without the stellar parameter cuts (see Table \ref{tab:summary_stats}). Lower panel: stars that pass the stellar parameter cuts. The vertical dotted lines in the upper panel indicate the stellar parameter cuts. 
    The color offset, $E(g-r)_\mathrm{star}-E(g-r)_\mathrm{map}$, is the difference between the per-star reddening measurement and the pixel-averaged value in the final calibrated map. Pixels with at least 16 stars are used for $E(g-r)_\mathrm{map}$. The solid and dashed black lines are the scatter from the 16th-84th percentile and RMS, respectively (the latter being larger due to deviation from zero mean and non-Gaussian tails of the distribution). The median and 16th and 84th percentiles are also shown. The gaps and excesses at regular spacing of the stellar parameters (which are also visible in Figure \ref{fig:calibration}) correspond to the grid lines in the PHOENIX stellar parameters and are a result of the inaccuracies in the interpolation by \texttt{rvspecfit}.}
    \label{fig:color_offset_trends_before}
\end{figure*}

The synthetic colors have systematic offsets that are strongly correlated with the stellar parameters and need to be removed before we can make the reddening measurements. Figure \ref{fig:color_offset_trends_before} shows the difference between the per-star reddening measurement in $g-r$ for the bright-time MWS stars in the South and the pixel average of the final (calibrated) map and how it varies with the stellar atmosphere parameters. The bright-time MWS stars are shown because they form the vast majority of the sample and their broad selection best illustrates the systematics trends. The offsets are most likely due to inaccurate stellar models, although inaccuracies in the assumed filter curves of the imaging may also contribute.

We model the systematic offset in the synthetic colors as a function of the stellar parameters:
\begin{equation}
(a-b)_{\mathrm{model},E=0} = (a-b)_{\mathrm{true},E=0} + f(p)
\end{equation}
where the offset $f$ is a function the set of stellar parameters $p$: $T_\mathrm{eff}$, log $g$, and [Fe/H]. (The correlation with [$\alpha$/Fe] is much weaker and we do not use it for calibration.) Our goal is to find this function and remove this offset.

The measured reddening of a star can written as the sum of the true reddening, the offset $f(p)$, and measurement error $\epsilon$:
\begin{equation}
\label{eqn:reddening_meas}
E(a-b)_\mathrm{meas} = E(a-b)_\mathrm{true} + f(p) + \epsilon
\end{equation}

In a nutshell, we obtain a noisy measurement of $f(p)$ for each star by comparing the measured reddening with the average reddening of nearby stars, and we reduce the noise by averaging over a large number of stars with similar stellar parameters. We first use SFD as a proxy for true reddening and obtain an initial estimate of $f(p)$. We then compute the $E(a-b)$ map using the calibrated reddening measurements and use this map as truth to re-evaluate $f(p)$. The second step is iterated until $f(p)$ converges. We detail the procedure below.

The initial offset is estimated by comparing the uncalibrated reddening measurement with the SFD-predicted reddening. This ties the absolute $E(B-V)$ zero point to SFD. Only low-extinction stars with $E(B-V)_\mathrm{SFD}<0.04$ are used in this initial step. For each star, we find its 400 nearest neighbors in the 3-dimensional stellar parameter space (with each dimension normalized by its median absolute deviation) for which we have an initial estimate of the offset $f(p)$. We take the mean offset of the 400 stars as the offset for the object. In other words, we avoid adopting a particular functional form of $f$ and instead compute it through the average reddening offsets of the nearest 400 stars in stellar parameter space. Besides anchoring the absolute zero point, another reason for implementing this initial calibration step using SFD is to reduce potential systematic errors from the model offsets. We later found that the results are the same without this SFD-dependent step: if we skip this step and instead use the uncalibrated reddening measurements in the iterative procedure described below, the resulting reddening map is essentially the same (except for an expected shift in the absolute zero point) --- the RMS difference in the $g-r$ reddening is only 0.1 mmag. Therefore we conclude that our reddening measurements do not depend on SFD except for the absolute zero point.

With the initial offset correction, we can now compute the average reddening for stars in each HEALPix pixel and create a HEALPix map with NSIDE of 128 (pixel size of 27.4$'$). Only pixels containing at least 16 stars are used. We repeat the procedure to calculate the per-star offset, now obtained by comparing the per-star reddening of each of the 400 nearest neighbors in stellar parameter space with the mean reddening of its corresponding pixel, and update the HEALPix map with the updated offset. The per-star offset calculation and map update are iterated for a total of 6 times (which are sufficient for the maps to converge) to obtain the final offset. The procedure is performed separately for $g-r$ and $r-z$.

While we could have performed the aforementioned procedure on the full sample, to avoid potential selection bias (see Section \ref{sec:selection_bias}), we instead perform the procedure on a subset of stars that forms the reference sample, which we use for the estimation of the offset $f(p)$ for the full sample. This reference sample is composed of MWS stars (MWS\_MAIN\_BLUE, MWS\_MAIN\_RED, MWS\_BROAD) and stars in the backup programs (BACKUP\_BRIGHT, BACKUP\_FAINT, GAIA\_STD\_FAINT and GAIA\_STD\_BRIGHT). They are selected because they are least affected by the selection bias. We also restrict to low-extinction regions by requiring $E(B-V)_\mathrm{SFD}<0.1$. Lastly, we randomly downsample the reference stars so there are no more than 50 stars in each HEALPix pixel with NSIDE of 128; this creates more uniform sampling (compared to the full sample which is dominated by stars at low Galactic latitudes). There are 1.51 million reference stars for $g-r$ and 1.27 million for $r-z$. The calibration procedure produces an intermediate reddening map from the calibration sample and the corresponding per-star offset estimates; these offsets are used for calibrating the full sample and they remain unchanged.

\begin{figure*}
    \centering
    \includegraphics[width=2.1\columnwidth]{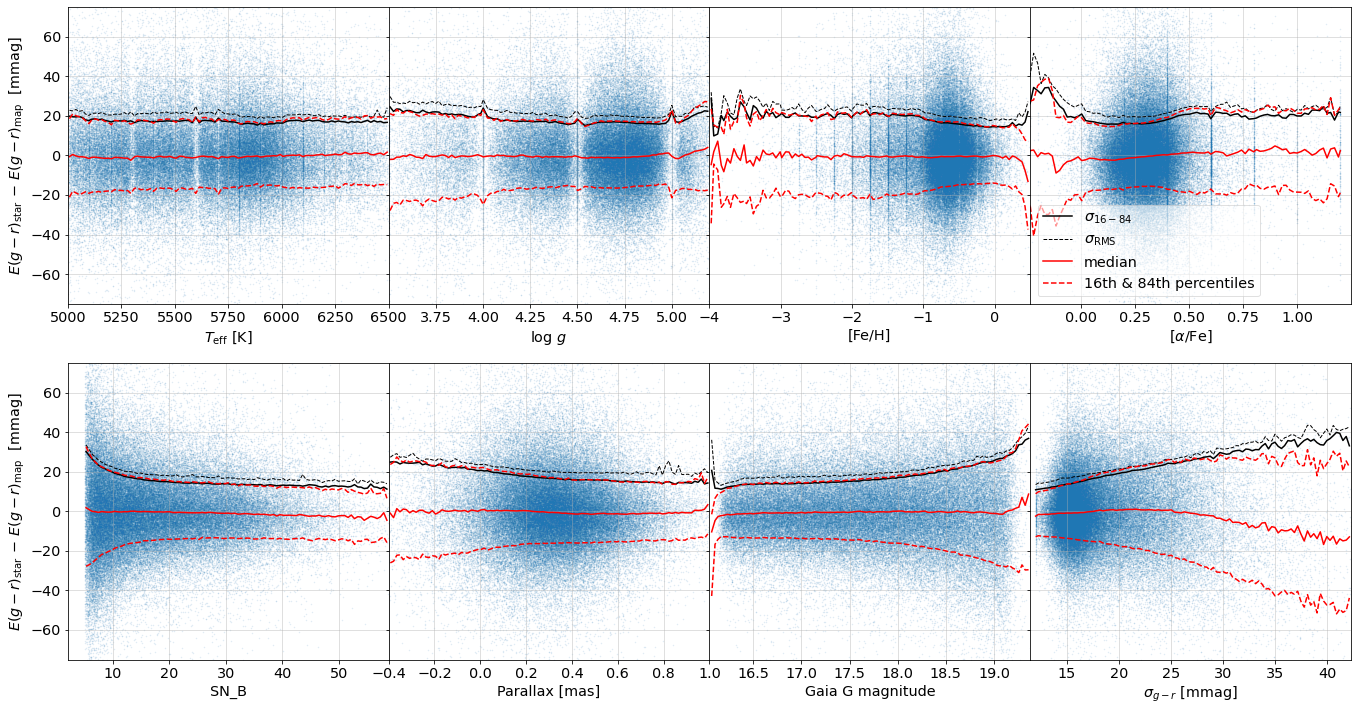}
    \caption{As Figure \ref{fig:color_offset_trends_before}, but showing the offsets after calibration. Also shown are the trends with the S/N in the blue spectrograph arm (SN\_B), parallax, Gaia $G$ magnitude, and $\sigma_{g-r}$ (the estimated error in $E(g-r)$; see Section \ref{sec:error_estimation}).}
    \label{fig:color_offset_trends_after}
\end{figure*}

\begin{figure*}
    \centering
    \includegraphics[width=2.0\columnwidth]{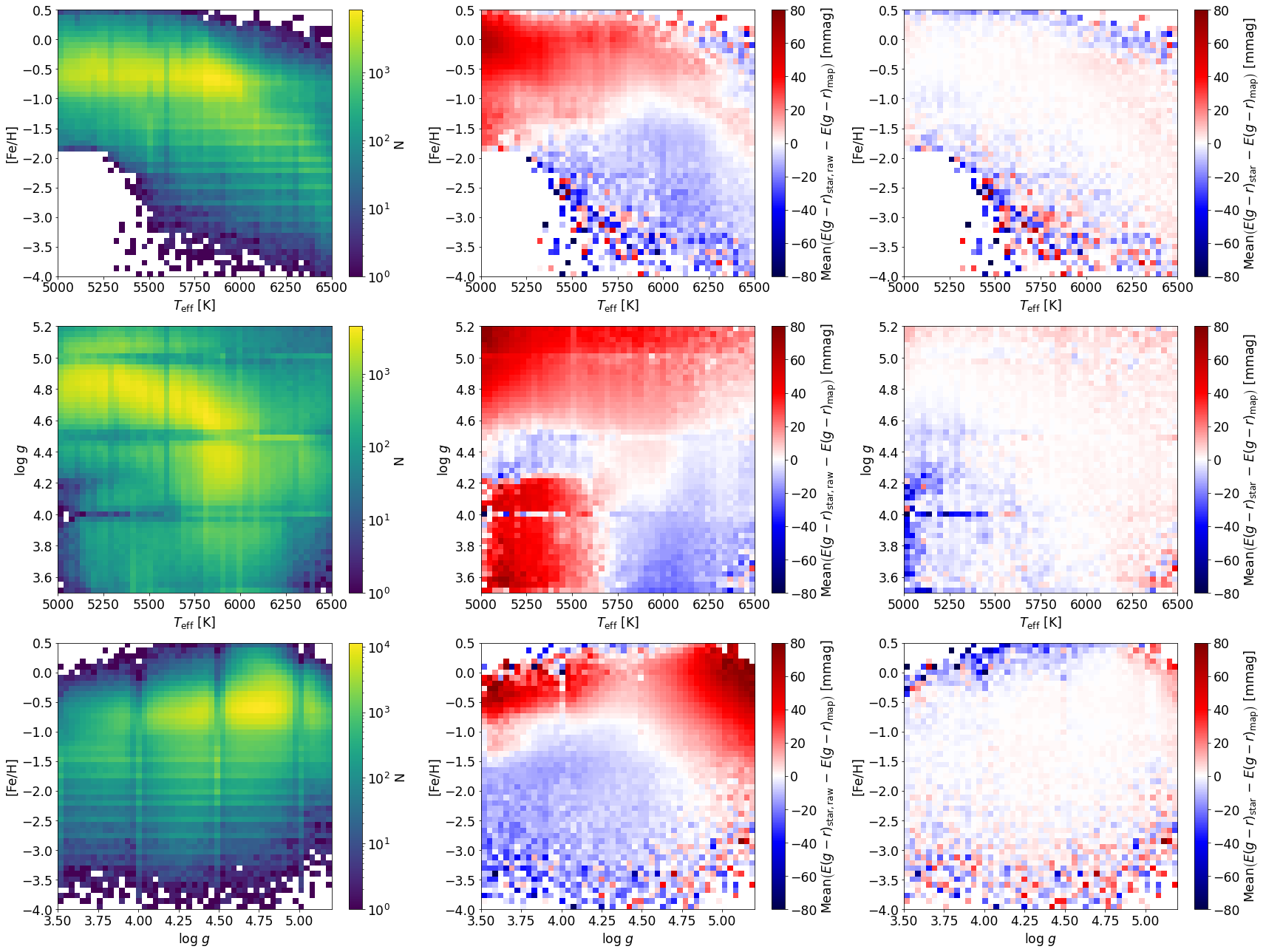}
    \caption{Similar to Figures \ref{fig:color_offset_trends_before} and \ref{fig:color_offset_trends_after} but showing the color offset in 2-D projections of combinations of $T_\mathrm{eff}$, [Fe/H] and log $g$. Left panels: the number of stars in each bin; middle panels: the mean of the uncalibrated (``raw'') $g-r$ offset; right panels: the mean offset after calibration.}
    \label{fig:calibration}
\end{figure*}

Finally, we calculate the offsets for the full sample with the same nearest-neighbor method. Rather than finding the 400 nearest neighbors in the full sample itself, we find them in the reference sample and use the mean offset of these 400 reference stars as the offset correction. 
Figure \ref{fig:color_offset_trends_after} shows the color offsets after calibration (compare to the offsets before calibration in Figure \ref{fig:color_offset_trends_before}), and it also shows trends with the blue-spectrograph S/N, parallax, Gaia $G$-band magnitude, and the estimated reddening error $\sigma_{a-b}$. Figure \ref{fig:calibration} shows the original offsets and residuals in 2-D projections of the stellar parameters.

\subsection{Error estimation of reddening measurements}
\label{sec:error_estimation}

Here we describe how we estimate the error in the calibrated reddening measurement for each star; these error estimates allow us to optimally create HEALPix maps using stars with different reddening errors and estimate the error in each pixel.

In the calibration procedure where we compute the mean offset $f(p)$ using the 400 nearest stars in stellar parameter space, we also record the standard deviation in the offset of these stars, $\sigma_{a-b,\mathrm{NN}}$. This quantity allows us to identify stars (or equivalently regions in the stellar parameter space) that are poorly modeled, and we remove these stars by imposing a maximum limit on $\sigma_{a-b,\mathrm{NN}}$: 30 mmag for $g-r$ and 35 mmag for $r-z$. We also use $\sigma_{a-b,\mathrm{NN}}$ as an intermediate quantity for estimating the error in the reddening measurement. While $\sigma_{a-b,\mathrm{NN}}$ can serve as the estimated error in the reddening measurement, it does not account for the variations in the spectroscopic S/N –– the same star with higher S/N typically has more accurate reddening measurements. Here we model the per-star reddening error $\sigma_{a-b}$ as a function of $\sigma_{a-b,\mathrm{NN}}$ and the blue-spectrograph S/N (SN\_B). The estimated per-star variance $\sigma_{a-b}^2$ is used in the inverse-variance weighting for the reddening maps.

\begin{figure*}
    \centering
    \includegraphics[width=1.8\columnwidth]{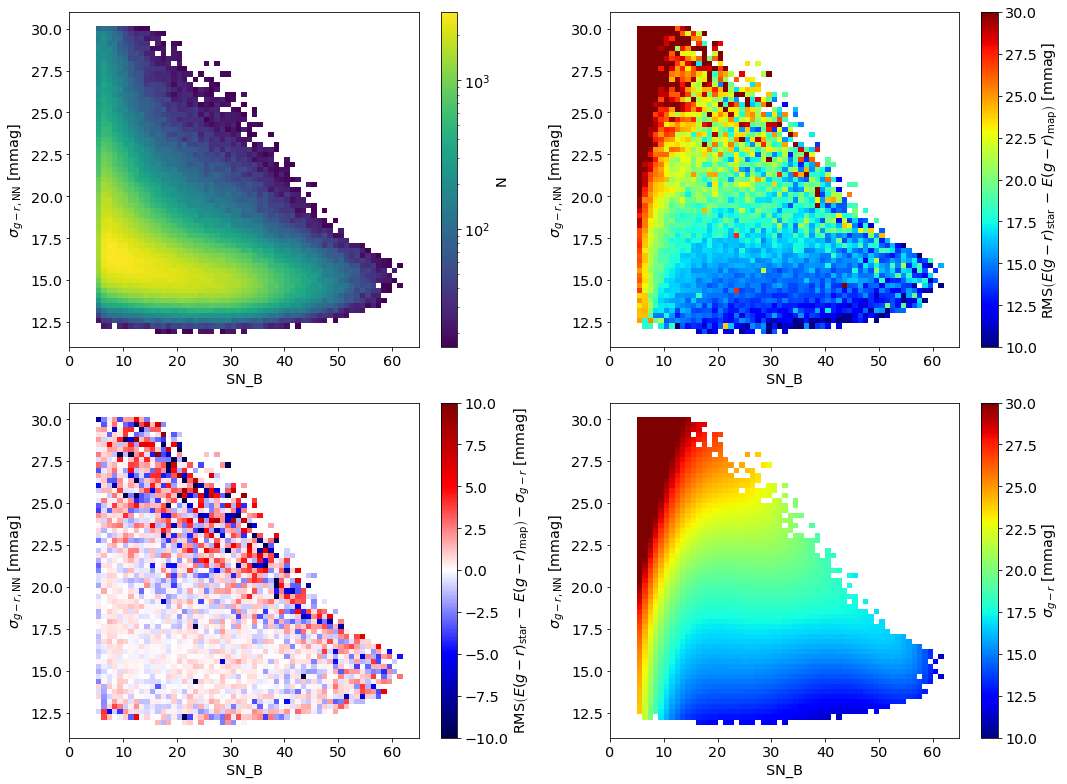}
    \caption{Error estimation for bright-time MWS stars in the South. Top left: the number of stars in each bin. Top right: the standard deviation in the $g-r$ offset in each bin. Bottom right: the error model based on polynomials of SN\_B and $\sigma_{g-r,\mathrm{NN}}$. Bottom left: the residual (i.e., the difference between the two right panels).}
    \label{fig:error_estimation}
\end{figure*}

We first compute the difference between the per-star reddening measurement $E(a-b)_\mathrm{star}$ and the average reddening of the corresponding HEALPix pixel (with NSIDE of 128) $E(a-b)_\mathrm{map}$ whose measurement error is much smaller than the individual measurement. We treat this difference as a result of the uncertainties in the reddening measurements. We only use stars in pixels with at least 32 stars and with $E(B-V)_\mathrm{SFD}<0.1$. We then divide the sample into 2-D bins of SN\_B and $\sigma_{a-b,\mathrm{NN}}$, keeping only bins with at least 16 stars, and for each bin compute $\sigma_{a-b}$, which is the RMS in $E(a-b)_\mathrm{star} - E(a-b)_\mathrm{map}$. We fit $\sigma_{a-b}$ as a function of SN\_B and $\sigma_{a-b,\mathrm{NN}}$ with 5th-order polynomials using the \texttt{statsmodels}' Robust Linear Models (\texttt{RLM}) routine, and compute the per-star reddening error using the polynomials. (We choose 5th-order polynomials as they are the lowest order polynomials that can adequately match the observed trends in diagnostic plots such as Figure \ref{fig:error_estimation}.) We impose an error floor of 12 mmag for both $E(g-r)$ and $E(r-z)$. Figure \ref{fig:error_estimation} shows the measured per-bin RMS errors, the model, and the residuals for the bright-time MWS stars in the South. For stars belonging to bins with fewer than 16 stars (typically in outlying regions in SN\_B and $\sigma_{a-b,\mathrm{NN}}$ where the polynomials may not be well behaved), they are assigned the error of their nearest neighbor in normalized SN\_B - $\sigma_{a-b,\mathrm{NN}}$ space. 

This procedure is performed separately for dark-time standard stars, bright-time standard stars, bright-time MWS stars, backup-time standard stars, and backup-time MWS stars. It is also done separately in the North and South (which have slight differences in their photometry), but we otherwise ignore variations in the photometric S/N. This is justified because the typical measurement error in the observed colors is $\sim$ 2 mmag which is much smaller than the typical error in the synthetic colors.

\subsection{Reddening maps}
\label{sec:reddening_maps}

We can now compute reddening HEALPix maps using the stellar reddening catalogs from the previous section.

First, we need to combine the North and South regions, which have been treated separately so far. While both regions have zero points that are matched to SFD, the resulting zero points in the two regions can differ because they are based on different regions of the SFD map that can have slightly different systematic offsets. We use the overlapping region between North and South to check for offsets in the reddening measurements, specifically using the overlapping pixels in the HEALPix maps. We find that relative zero-point offsets (after correcting for the different extinction coefficients), in $E(a-b)_\mathrm{North}-E(a-b)_\mathrm{South}$, are $-9.5$ mmag for $E(g-r)$ and $-5.2$ mmag for $E(r-z)$. We artificially remove this offset from the reddening values in the North such that their zero point matches the South. We then normalize the Northern reddening measurements so that they match the Southern measurements. E.g., for $g-r$ reddening, we multiply the Northern measurements by $(R_{g,\mathrm{south}} - R_{r,\mathrm{south}})/(R_{g,\mathrm{north}} - R_{r,\mathrm{north}})$. The reddening maps that we produce are thus entirely in the DECam system.

To combine the two regions into a single catalog, we only keep stars with DEC$>32.375^{\circ}$ in the Northern catalog, and stars with DEC$\leq 32.375^{\circ}$ or in the South Galactic Cap (SGC) in the Southern catalog. This produces a combined North+South catalog with unique stars with $E(g-r)$ measurements, and a separate catalog with $E(r-z)$ measurements. As mentioned earlier, the two catalogs have slightly different sets of stars due to the different quality cuts.

\begin{figure*}
    \centering
    \includegraphics[width=2.1\columnwidth]{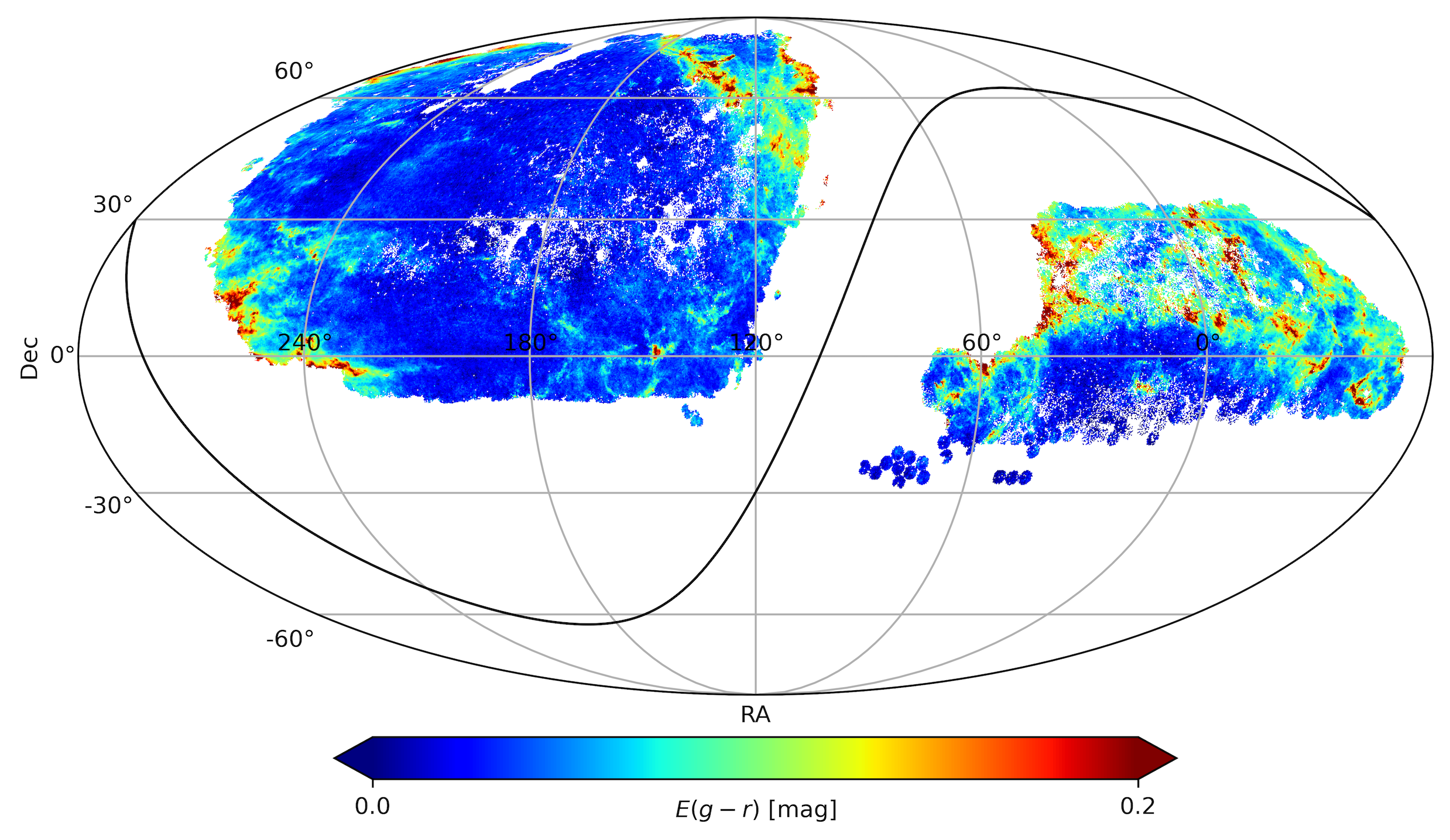}
    \caption{DESI $g-r$ reddening map. The map has a resolution of NSIDE=256 (13.7$'$ pixel size).}
    \label{fig:egr_map}
\end{figure*}

\begin{figure*}
    \centering
    \includegraphics[width=2.1\columnwidth]{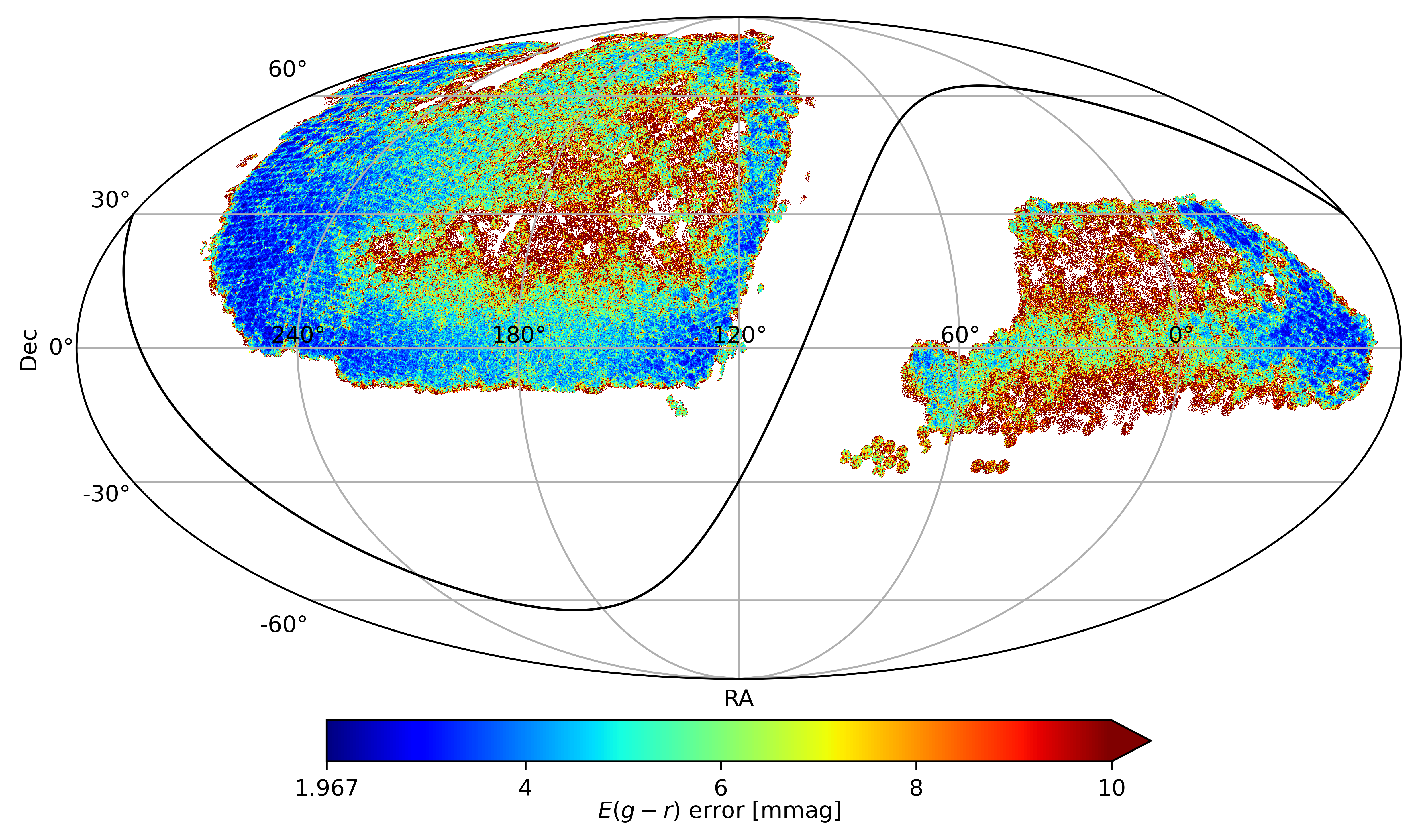}
    \caption{The estimated per-pixel (NSIDE=256) error in the  $g-r$ reddening measurement.}
    \label{fig:egr_error_map}
\end{figure*}

We use the combined stellar catalog to create the HEALPix map. For each pixel, we compute the inverse variance-weighted average of the per-star reddening measurements. The estimated inverse variance of the reddening of each pixel is the sum of the inverse variance of its stars. We create HEALPix maps with 3 different resolutions: NSIDE of 128, 256, and 512 (corresponding to pixel sizes of 27.5$'$, 13.7$'$ and 6.9$'$). Figure \ref{fig:egr_map} shows the $g-r$ reddening map (with NSIDE of 256), and Figure \ref{fig:egr_error_map} shows the map of the estimated errors.

The discrete (and in some areas sparse) nature of the stellar sample means that not all HEALPix pixels will contain stars, and the fraction of ``missing'' pixels increases with resolution. The missing pixels artificially increase the angular power $C_\ell$ at small scales, especially at high resolutions, and this complicates the comparison with the SFD map. Here we interpolate over (or ``inpaint'') the missing pixels using reddening measurements of nearby ones. This is done by iteratively convolving the map with a Gaussian kernel to fill in the missing pixels; the values of the valid pixels are unchanged. See Appendix \ref{sec:filling} for details. The interpolation assumes that the spatial variation is sufficiently smooth, but it may not be strictly true. Therefore the interpolated reddening estimates are less reliable than the actual measurements, and we advise caution when using them. The reddening maps shown throughout this paper have no interpolation, and the interpolated maps are only used for computing the power spectrum $C_\ell$ in Figure \ref{fig:c_ell}.

\section{Discussion}
\label{sec:discussion}

\subsection{Comparison with SFD}
\label{sec:sfd_comparison}

\begin{figure*}
    \centering
    \includegraphics[width=2.1\columnwidth]{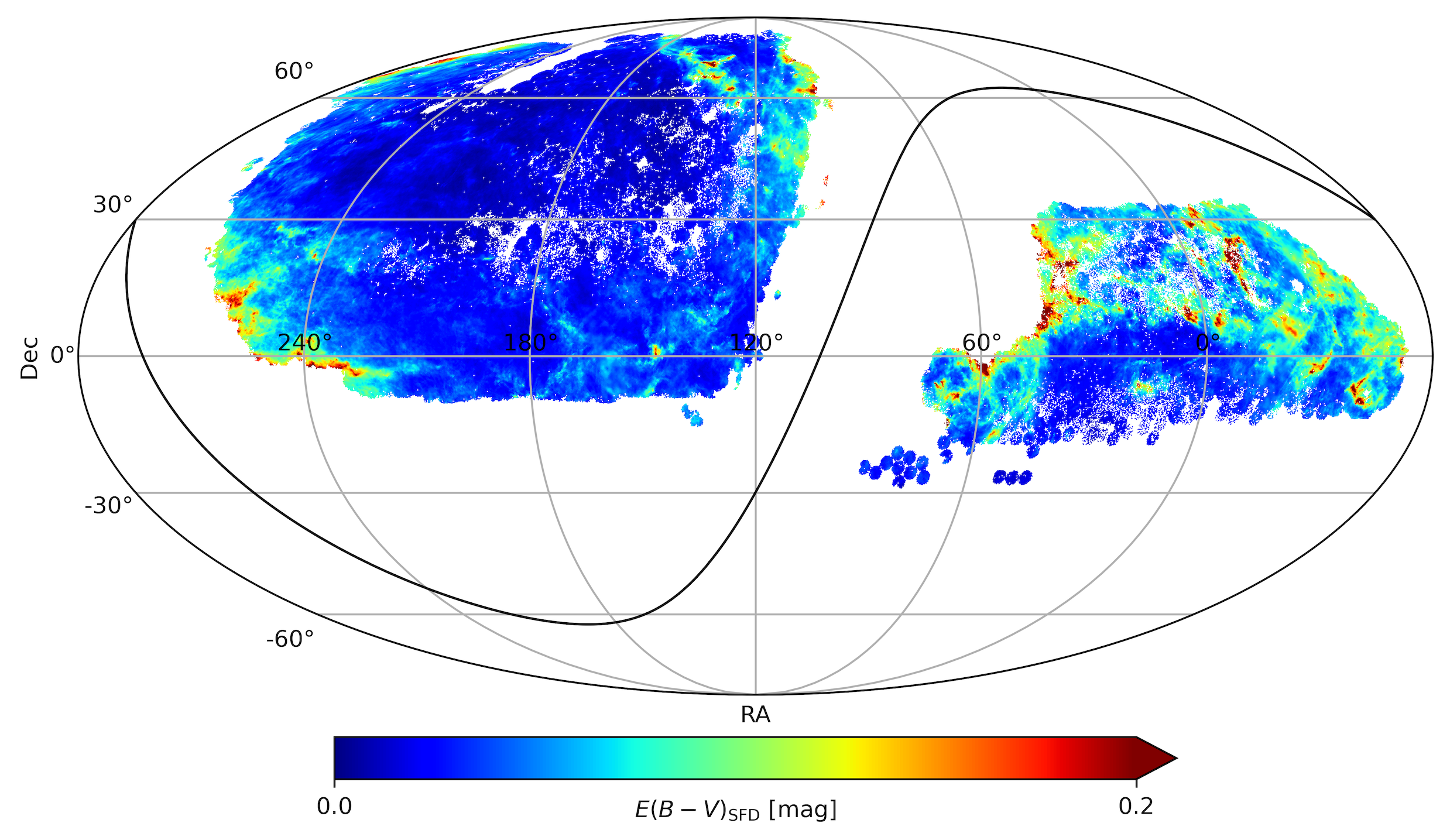}
    \includegraphics[width=2.1\columnwidth]{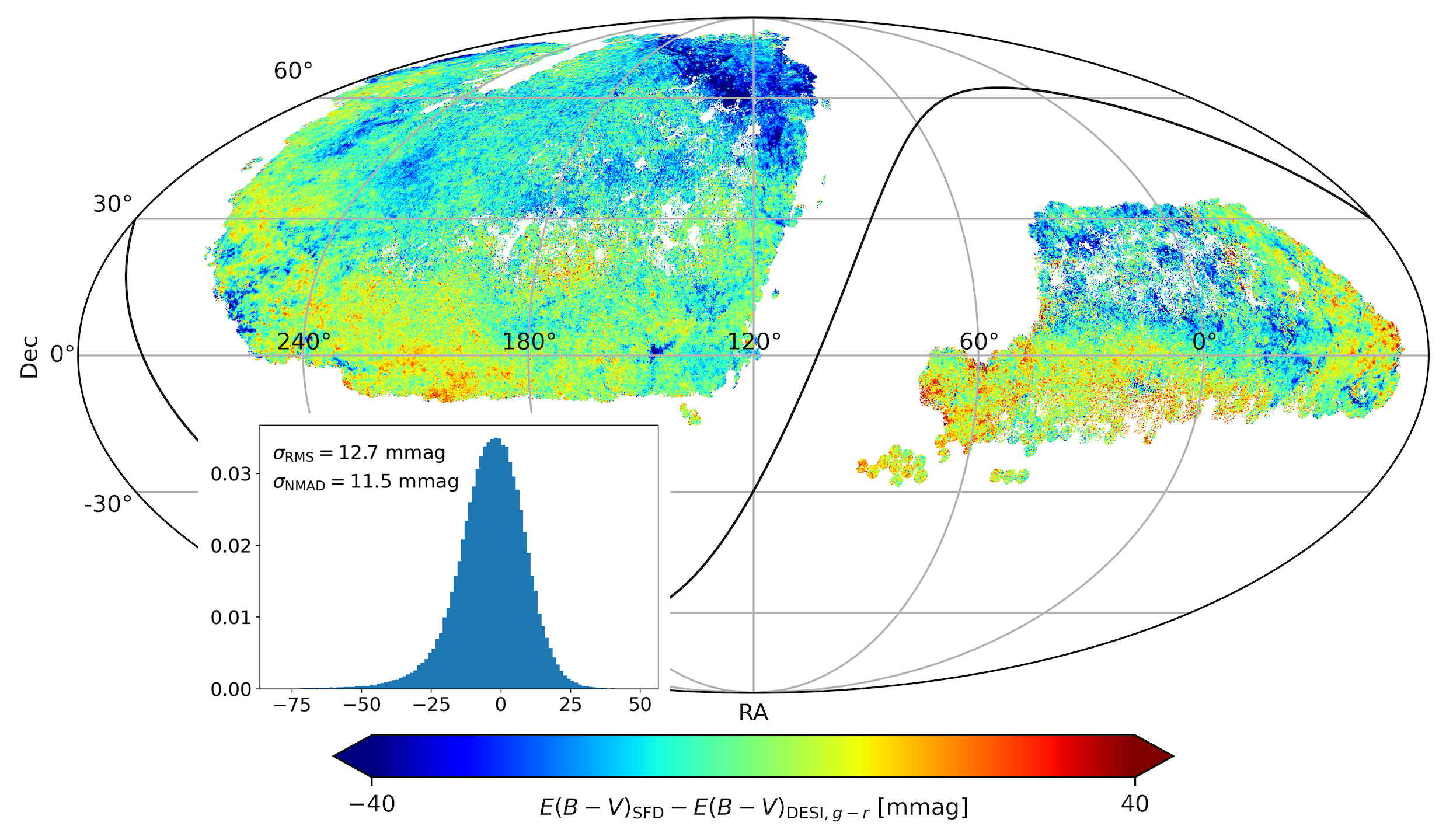}
    \caption{Top: the SFD $E(B-V)$ map, showing only pixels with valid measurements in the DESI $E(g-r)$ map. The SFD $E(B-V)$ values are sampled at the same location as the stars in the DESI $g-r$ sample. Bottom: the difference between SFD and the DESI-derived $E(B-V)$ (which is $E(g-r)/1.049$). The bottom panel inset shows the histogram of the difference and some statistics. $\sigma_\mathrm{NMAD}$ is the normalized median absolute deviation. All valid pixels are shown in the maps, but only pixels with DESI $E(B-V)$ error less than 9 mmag (which are 80\% of the pixels; the median error is 5 mmag) are used for the histogram and statistics in the inset, so that they are not dominated by noisy measurements. Both maps have NSIDE=256.}
    \label{fig:desi_vs_sfd}
\end{figure*}

\begin{figure*}
    \centering
    \includegraphics[width=2.1\columnwidth]{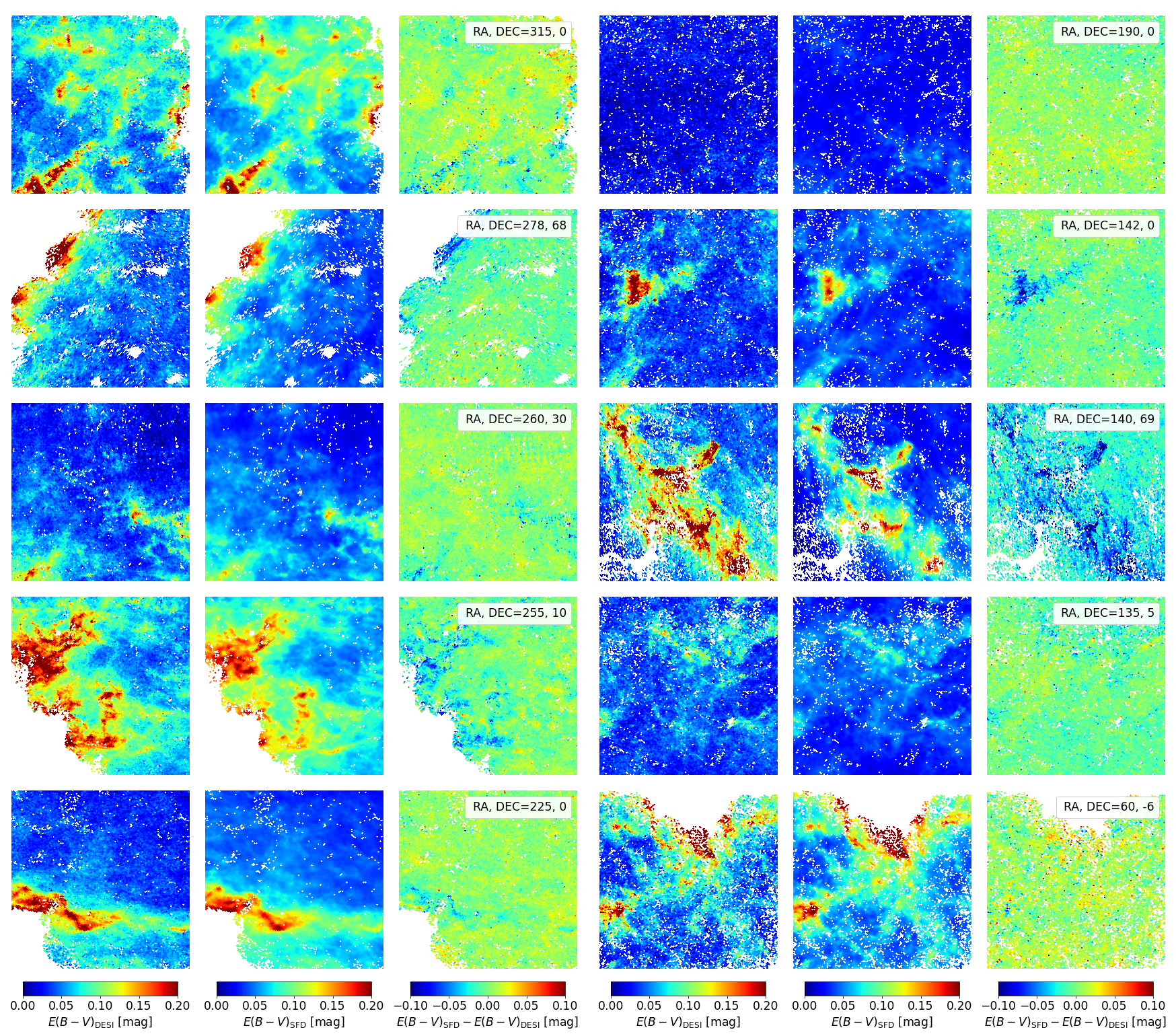}
    \caption{Zoomed-in views of the $E(B-V)$ map derived from DESI $E(g-r)$ (first and fourth columns), the SFD map (second and fifth columns), and their difference (third and sixth columns). Here only pixels with valid $E(g-r)$ measurements (i.e., containing at least one star) are shown for both DESI and SFD. Maps with NSIDE=512 (6.9$'$ pixel size) are shown. Each image is 15 degrees by 15 degrees.}
    \label{fig:desi_vs_sfd_zoom}
\end{figure*}

The direct reddening measurement from DESI allows us to perform a detailed comparison with the SFD $E(B-V)$ map. Figures \ref{fig:desi_vs_sfd} and \ref{fig:desi_vs_sfd_zoom} show the comparison between the SFD and the $E(B-V)$ inferred from DESI $g-r$ reddening. Our map shows a very strong correlation with SFD; we are clearly measuring real extinction in the Milky Way. However, also evident in the two figures are substantial differences, highlighting the value of direct extinction measurements as opposed to relying on maps of thermal dust emission to predict optical extinction.

Immediately visible are areas with large differences between the two maps. E.g., SFD underpredicts $E(B-V)$ by as much as 80 mmag in the north-western corner of the North Galactic Cap (NGC) (this difference was also visible in, e.g., \citealt{schlafly_measuring_2011}), and the differences show large spatial variations in the South Galactic Cap (SGC). The DESI map's high resolution also reveals many smaller features in the difference map, such as the compact area at RA=$146^{\circ}$, DEC=$1^{\circ}$ and small-scale variations in the higher-extinction regions.

\begin{figure}
    \centering
    \includegraphics[width=1.\columnwidth]{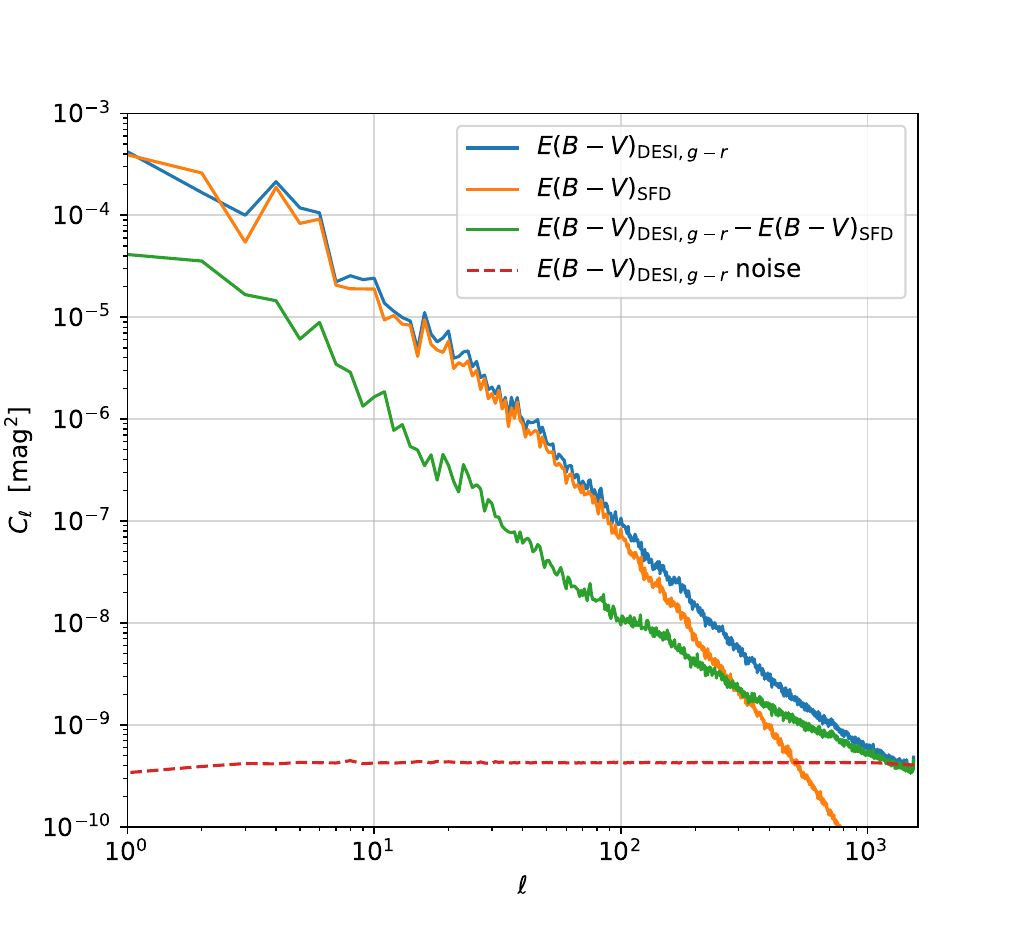}
    \caption{Angular power spectra of the DESI and SFD $E(B-V)$ maps, and their difference. The red dashed line is the DESI noise spectrum that assumes the estimated per-pixel errors. Maps with NSIDE of 512 are used. See text for more details.}
    \label{fig:c_ell}
\end{figure}

Figure \ref{fig:c_ell} shows the angular power spectra of the DESI and SFD maps and their difference. To compute $C_\ell$, we use the inpainted DESI $E(B-V)$ map with NSIDE=512 (with $E(B-V)$ in units of magnitude) with the median $E(B-V)$ subtracted off. For SFD we only use the pixels where DESI measurement is available and inpaint the missing pixels the same way that the DESI map is inpainted; as such the $C_\ell$'s are from the same sky area with the same interpolation scheme. We use the \texttt{sphtfunc.anafast} routine from \texttt{healpy} and we correct for the missing sky fraction by rescaling $C_\ell$ by a factor of $1/f_\mathrm{sky}$ (where $f_\mathrm{sky}$ is the fraction of ``good'' pixels defined in appendix \ref{sec:filling}).

The two maps have a $\sim$ 10\% difference at larger scales ($\ell<\sim 100$). As we will show in Section \ref{sec:elg_selection}, this difference can be mostly attributed to systematic errors in the SFD map. The DESI map also has significantly more power at smaller scales ($\ell>\sim 100$), and this difference cannot be explained by the uncorrelated per-pixel noise that we expect is in the DESI map, as the noise only starts to dominate the signal at $\ell>\sim 800$ --- the red dashed line in Figure \ref{fig:c_ell} shows the expected noise power spectrum of the DESI map, which is estimated by generating maps of random Gaussian values with the actual per-pixel error $\sigma_{g-r}$. The difference at smaller scales also cannot be explained by the resolution limit of the SFD map, as its FWHM of 6.1$'$ roughly corresponds to $\ell=1800$. This difference suggests that there may be small-scale variations in the dust extinction that are captured by the DESI map but not by the SFD map.

\subsection{Validating the maps using galaxies as back light}
\label{sec:elg_selection}

The observations of galaxies are also affected by the dust of our galaxy. While it is more difficult to determine the intrinsic spectra of galaxies and estimate the galaxy reddening directly (although possible, e.g., see \citealt{peek_correction_2010}), we can exploit the fact that the distribution of galaxies is isotropic at large scales, and use galaxies as a uniform back light to test the reddening maps. It is worth noting that the pioneering work of \cite{bursteinReddeningsDerivedGalaxy1982} made use of galaxy counts to produce a reddening map that was widely used before the SFD map became available. Here we use the DESI emission line galaxy (ELG) sample \citep{raichoor_target_2023} as the back light sample to test the DESI and SFD maps.

The DESI ELGs are particularly sensitive to extinction due to its reliance on $g$-band magnitude and $g-r$ color: a $+10$ mmag ($-10$ mmag) change in $E(B-V)$ causes a change in the overall target density of ELG\_LOP (the primary ELG target type used for cosmology; hereafter ELG for simplicity) by $+11.0\%$ ($-10.1\%$), compared to $-1.4\%$ ($+1.3\%$) for the DESI luminous red galaxy targets \citep{zhou_target_2023}.

\begin{figure}
    \centering
    \includegraphics[width=1.\columnwidth]{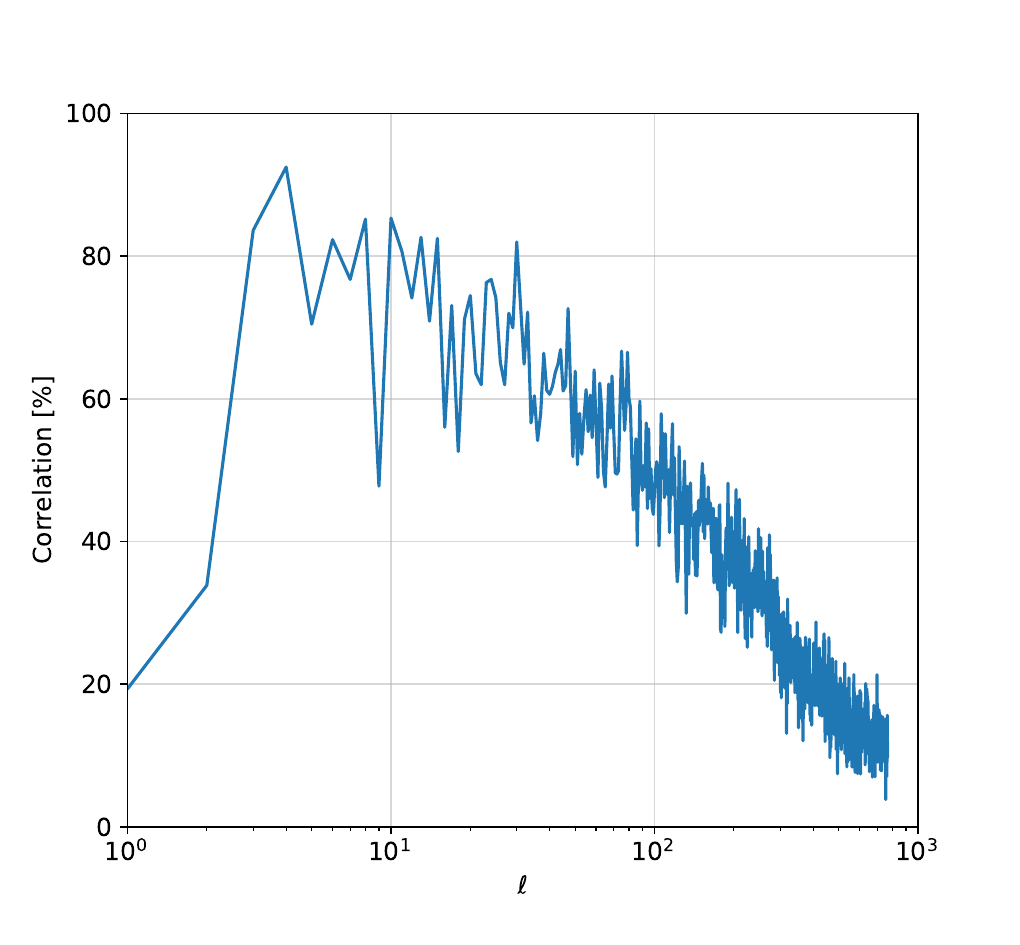}
    \caption{The correlation ($C_\ell^{12}/\sqrt{C_\ell^{11} C_\ell^{22}}$) between map 1: the DESI ELG target density, and map 2: the difference in $E(B-V)$ between the SFD and DESI maps, i.e., $E(B-V)_\mathrm{SFD} - E(B-V)_{\mathrm{DESI},g-r}$. Maps with NSIDE of 256 are used.}
    \label{fig:elgxdebv}
\end{figure}

The DESI targets were selected using SFD for extinction correction and are thus affected by the systematic errors in the SFD map, as any inaccuracy in the extinction map can produce an artificial excess or deficit in the projected galaxy density. Density modulations matching $E(B-V)_\mathrm{SFD} - E(B-V)_{\mathrm{DESI},g-r}$ (lower panel of Figure \ref{fig:desi_vs_sfd}) are seen in the ELG density map (see Figure 6 of \citealt{raichoor_target_2023}). Figure \ref{fig:elgxdebv} shows the correlation between the DESI ELG target density and the $E(B-V)$ difference map --- they are correlated at the $\sim 80\%$ level at the largest scales! The ELG selection's sensitivity to (errors in) extinction presents a significant complication for DESI cosmology analyses, and this was the primary motivation for creating this new reddening map.

\begin{figure*}
    \centering
    \includegraphics[width=1.45\columnwidth]{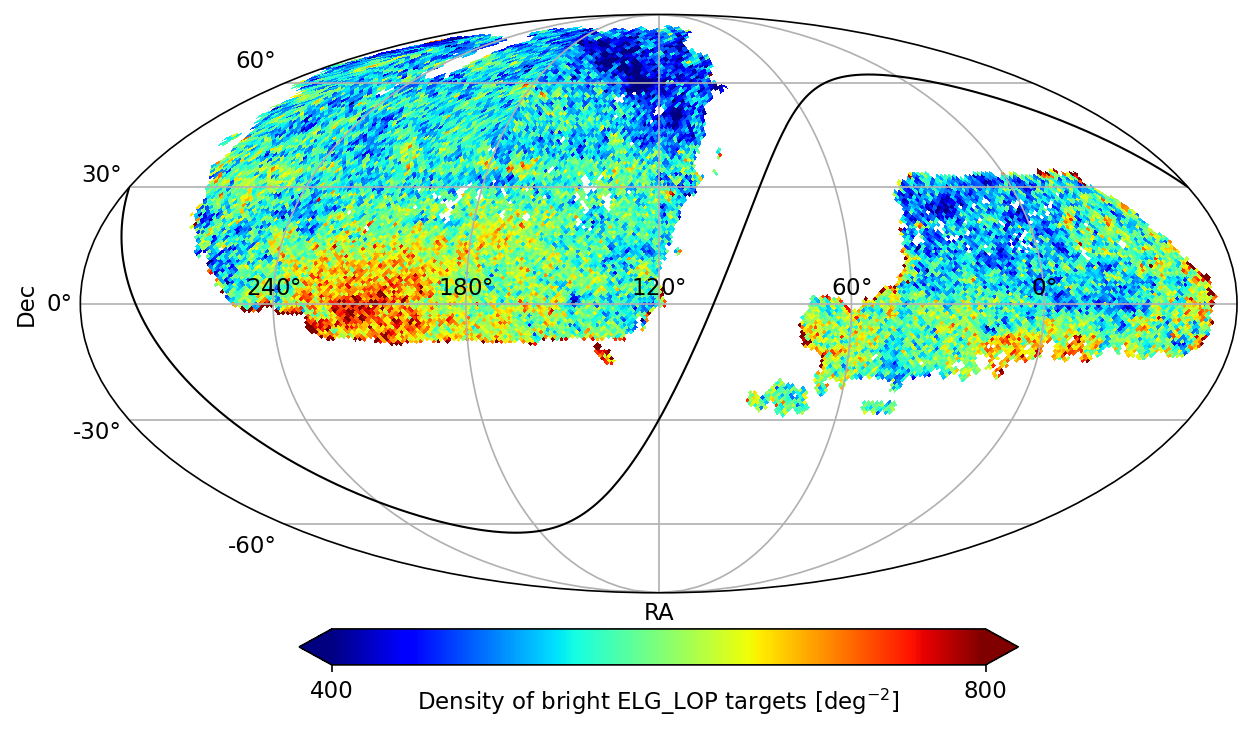}
    \includegraphics[width=1.45\columnwidth]{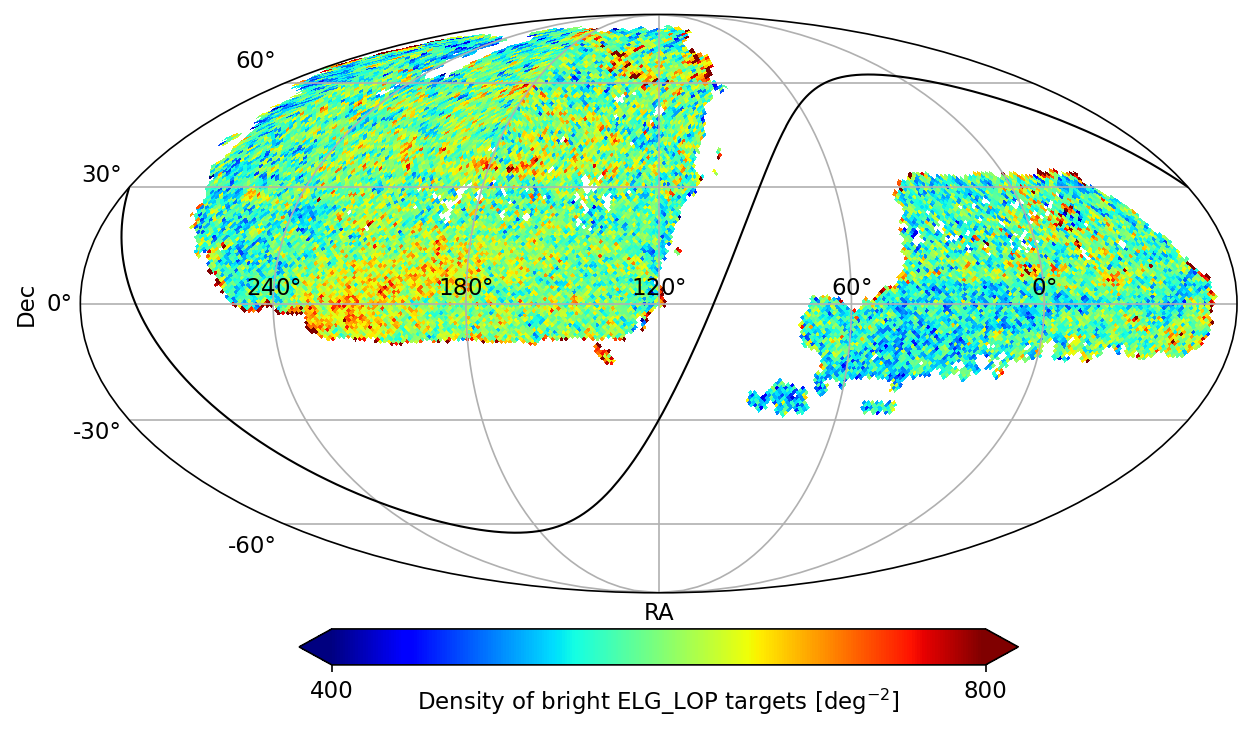}
    \caption{The density of bright ELG targets (see Section \ref{sec:elg_selection}) selected using different reddening maps for extinction correction. The top panel uses the SFD map and the bottom panel uses the DESI $E(g-r)$ map. Only pixels with DESI reddening $E(g-r)$ errors less than 6 mmag are shown. Both maps have NSIDE=64.}
    \label{fig:elg_density_maps}
\end{figure*}

To test the DESI reddening map, we reselect DESI ELG targets using DESI $E(g-r)$ for extinction correction and compare the resulting density map with the SFD-based selection. To reduce the effects due to other imaging systematics such as depth and seeing variations that would otherwise complicate this comparison, we apply a brighter magnitude limit: we implement a $g$-band fiber magnitude limit of $<23.6$ (0.5 mag brighter than the original selection) but otherwise keep all ELG\_LOP cuts unchanged. Figure \ref{fig:elg_density_maps} shows the resulting density maps using SFD and DESI $E(B-V)$. The DESI reddening-based ELGs are significantly more uniform, and systematics unrelated to extinction (including depth variations, stellar contamination, and zero point errors) now dominate the density variations. For more ELG-related discussions, see \cite{alberto_elg_imaging_sys} which describes the modeling of imaging systematics in the Y1 spectroscopic ELGs using the DESI reddening map.

\subsection{Validating the extinction curve with reddening color-color diagram}
\label{sec:color-color}

\begin{figure}
    \centering
    \includegraphics[width=0.98\columnwidth]{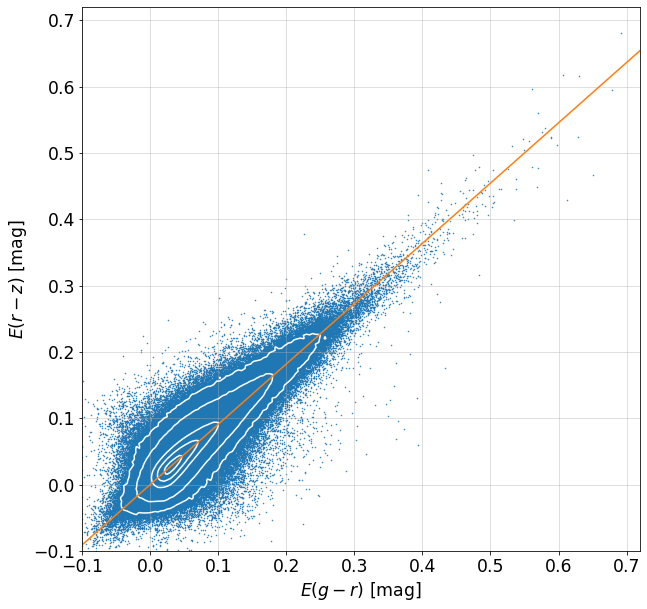}
    \caption{Reddening color-color diagram. Each point is the reddening measurement of a star. The contours enclose 10\%, 40\%, 68\%, 95\% and 99\% of the sample. The orange line, which has slope of $(R_r-R_z)/(R_g-R_r)$, is the expected trend for the assumed extinction coefficients $R_x\equiv A_x/E(B-V)$ based on the $R_\mathrm{V}=3.1$ \cite{fitzpatrick_correcting_1999} extinction curve.}
    \label{fig:color-color}
\end{figure}

Here we validate the assumed extinction curve by comparing the reddening in the $g-r$ and $r-z$ colors. Figure \ref{fig:color-color} shows the reddening color-color diagram, $E(g-r)$ vs $E(r-z)$, for the entire stellar sample, and we find that on average the reddening in the two colors is consistent with the $R_\mathrm{V}=3.1$ \cite{fitzpatrick_correcting_1999} extinction curve, reaffirming the finding in \cite{schlafly_measuring_2011}.

Previous studies, such as \cite{schlaflyOpticalinfraredExtinctionCurve2016} and \cite{zhangThreedimensionalMapsInterstellar2025}, have measured significant spatial variations in $R_\mathrm{V}$, but they are limited to high-extinction regions at low Galactic latitudes where $R_\mathrm{V}$ can be more easily measured. Here we check for spatial deviations in the reddening curve by comparing the reddening in the two colors, $g-r$ and $r-z$; see Figure \ref{fig:ebv_gr_vs_rz}. However, this difference is degenerate with zero point offsets in the $grz$ bands due to photometric calibration errors in the LS DR9 imaging data, and we find that the measured difference can be mostly explained with the zero point offsets measured with Gaia. We discuss the zero point offsets in Section \ref{sec:zp_systematics}. The zero point uncertainties make it difficult to interpret the residual differences or make quantitative assessments. Nevertheless, we do not see deviations from the universal extinction curve that are significant compared to the photometric calibration errors within our survey footprint.

\section{Sources of potential systematic errors}
\label{sec:systematics}

\subsection{Photometric zero point uncertainties}
\label{sec:zp_systematics}

One major source of systematic error in any flux measurement is the uncertainties in the photometric zero point calibration. The LS DR9 photometric zero points are tied to Pan-STARRS1 \citep{chambers_panstarrs1_2019} and thus inherit its zero-point systematics. For the reddening measurements, photometric calibration errors can be thought of as a component in the measurement error $\epsilon$ in equation \ref{eqn:reddening_meas}. However, because they are a systematic error that varies spatially, they cannot be averaged out in the reddening map, and the reddening measurements are completely degenerate with zero point offsets.

Indeed, the photometric calibration errors are by far the largest source of error in our reddening measurements at larger scales. They may also affect smaller scales (e.g., due to incorrect flat-fielding), although it's more difficult to assess such effects with data that is currently available. And, unlike the measurement errors (which we estimate in Section \ref{sec:error_estimation}) due to photometric/spectroscopic noise and modeling uncertainties, the photometric calibration error is a systematic error that cannot be reduced by increasing the density of the spectroscopic sample.

\begin{figure*}
    \centering
    \includegraphics[width=1.45\columnwidth]{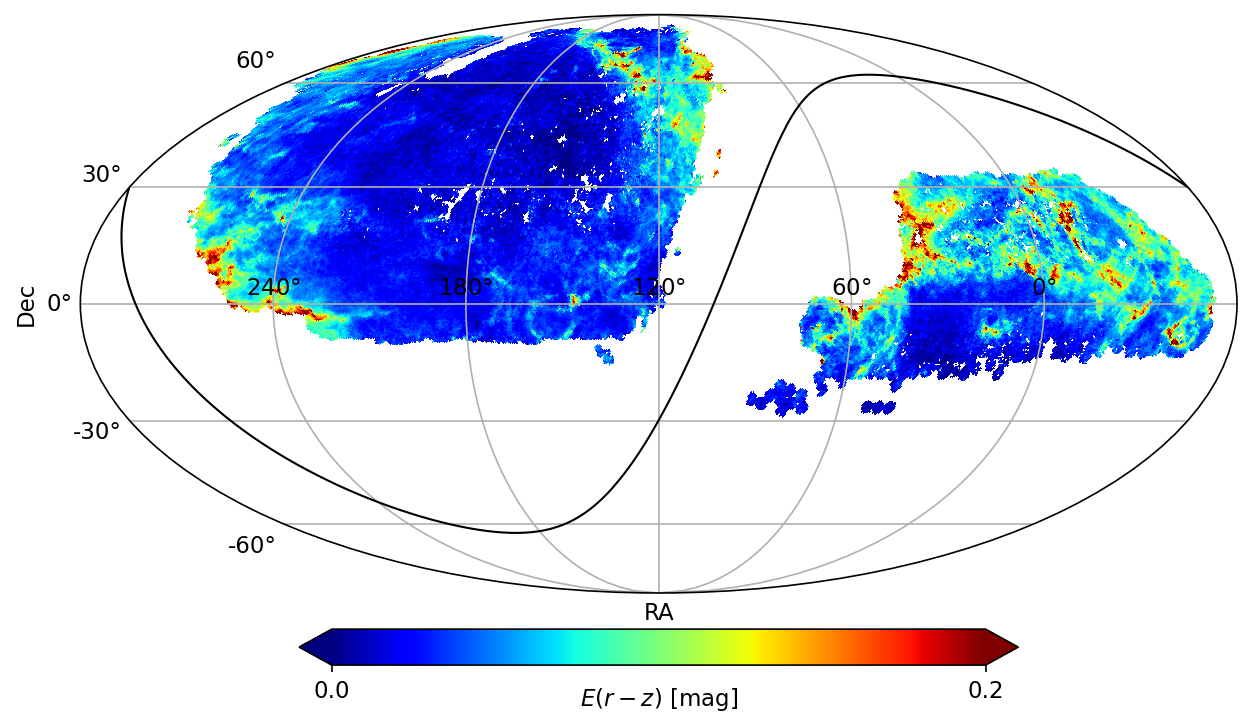}
    \includegraphics[width=1.45\columnwidth]{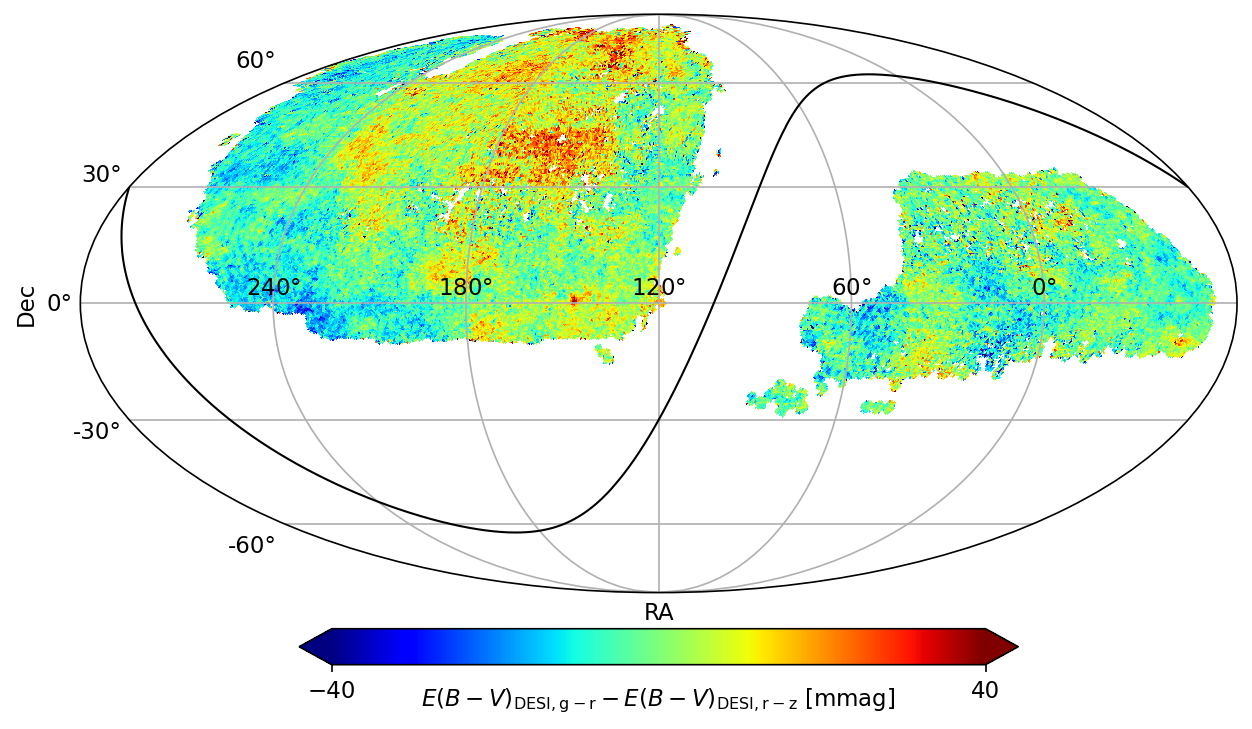}
    \includegraphics[width=1.45\columnwidth]{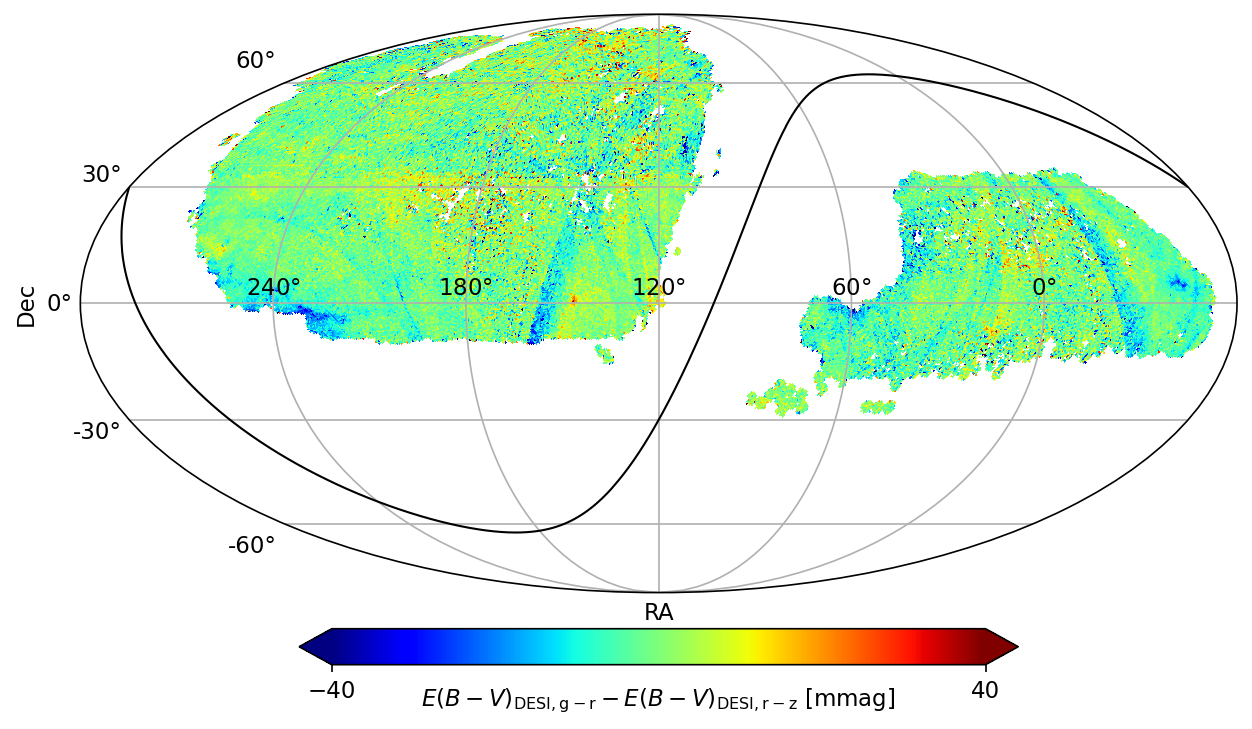}
    \caption{Top panel: the $r-z$ reddening map. Middle panel: the difference between the $E(B-V)$ map from $g-r$ and the $E(B-V)$ map from $r-z$. Bottom panel: the difference map after removing the zero point offsets measured from Gaia spectrophotometry. Most of the features in the difference map are due to zero point systematics as they disappear after the Gaia correction. The stripes in the ``corrected'' difference map are due to systematics in the Gaia spectrophotometry. All three maps have NSIDE=128. Note that the bottom panel is only shown for illustration, and we do not apply the Gaia correction to our maps and catalogs.}
    \label{fig:ebv_gr_vs_rz}
\end{figure*}

The effects of zero point systematics are apparent when comparing the $E(B-V)$ maps inferred from $g-r$ and from $r-z$. If the extinction curve is the same everywhere, we would expect the two maps to be identical. The middle panel of Figure \ref{fig:ebv_gr_vs_rz} shows the difference between the two maps; the bottom panel shows the difference in the $g-r$ and $r-z$ reddening maps when we ``correct'' for the zero-point offsets using the zero point offsets maps from \cite{zhou_comparing_2023} that uses Gaia spectrophotometry as the reference photometry. This zero-point ``correction'' removes most of the features in the $E(B-V)$ difference but also introduces the Gaia systematics (e.g., the scan patterns along the declination direction). To avoid contamination by the systematics in the Gaia spectrophotometry, we do not apply this ``correction'' on the reddening measurements in our data products.

As reported in \cite{zhou_comparing_2023}, the zero point errors of LS DR9 are 4.7, 3.7, 4.4 mmag in DECam $grz$ bands, respectively. The resulting errors in our reddening measurements, roughly 6 mmag in both $E(g-r)$ and $E(r-z)$, are thus not negligible given our overall measurement errors. However, we should note that the zero point errors only contribute a fraction of the difference with SFD, and the features in the bottom panel of Figure \ref{fig:desi_vs_sfd} are mostly due to systematic errors in the SFD map. The angular power of the difference map ($E(B-V)_\mathrm{DESI} - E(B-V)_\mathrm{SFD}$) is reduced by only $5-20\%$ at large scales ($\ell<\sim300$) when the zero point ``correction'' is applied.

Note that when the reddening measurements and sources of interest (to be dereddened) are based on the same imaging data (e.g., using DESI imaging for both reddening measurements and ELG selection), the dereddened colors of these sources do not contain spatially-varying calibration systematics. This is because both the observed color (of the sources of interest) and the reddening measurement contain the same photometric zero-point offsets, and they cancel out in the dereddened colors. Also note that 1) the extinction-corrected \textit{magnitudes} still contain the spatially-varying calibration errors, and 2) the extinction-corrected colors still contain the \textit{absolute} zero point errors.

\subsection{Photometric selection bias}
\label{sec:selection_bias}

The stellar sample used in this work was not originally designed for measuring reddening, and its selection was not optimized for this purpose. One complication arises from the fact that the same $g-r$ and $r-z$ colors are used for both sample selection and reddening measurement.

Because the observed colors have non-zero errors, the color cuts can produce a bias for the reddening measurement. For example, the standard stars are only selected if the SFD-extinction-corrected $g-r$ color is bluer than 0.35, so a star near the selection boundary will be more likely to be selected if the $g-r$ measurement randomly scatters in the bluer direction than in the redder direction. The result is an error distribution that is no longer Gaussian but tilted towards the blue, effectively causing a blue bias. Almost all ($\gtrsim 99\%$) of the stars in our sample have measurement errors in $g-r$ and $r-z$ of less than 6 mmag (with a mean of 2 mmag) based on the photometric errors reported by \texttt{tractor} (the algorithm that performs source extraction on the images; \citealt{Lang16}), so this is a relatively small effect. There is no easy way to completely remove the selection bias from the sample, and we do not attempt to correct this bias in our reddening measurements. Nonetheless, it informs our choices in calibrating the synthetic colors in Section \ref{sec:calibration}.

The standard stars (except for those in backup tiles, as they are targeted using Gaia photometry) are most affected by this selection bias because there are a significant number of stars near the color cut boundaries (in $g-r$ and $r-z$). Because the colors are strongly anti-correlated with $T_\mathrm{eff}$, if these stars are used in the calibration of the synthetic color, the calibration would attempt to correct for the bias at low $T_\mathrm{eff}$, and this would introduce a bias to the other stars that are not affected by this bias. For this reason, the standard stars are excluded from the calibration reference sample (see Section \ref{sec:calibration}).

The MWS stars are selected using the $r$-band magnitude for the magnitude limits and the $g-r$ color to distinguish between different subsamples (which have different observing priorities). However, the MWS stars are much less affected by the selection bias because a much smaller fraction of them are near the selection boundaries compared to the standard stars. For this reason, and because the MWS stars are a large sample that spans the entire stellar parameter space, we include the MWS stars as the reference sample. We also include stars from the backup program in the reference sample because they are selected using Gaia photometry and are thus not affected by the selection bias.

\subsection{Dependence of the synthetic colors on the amount of extinction}
\label{sec:model_extinction_bias}

For the reddening measurements, we have assumed that the synthetic colors from the best-fit \texttt{rvspecfit} models do not depend on the actual amount of extinction of the observed stars. To verify this assumption, we perform the following test. We randomly select a total of 99 stars that sample the different observing programs and target classes, and artificially add (or remove) extinction by multiplying (or dividing) the observed spectra with the additional dust extinction (assuming the \citealt{fitzpatrick_correcting_1999} extinction curve). For each star, we generate new ``observed'' spectra with ``added''/``removed'' extinction $\delta E(B-V)$ in a 0.05 mag-spaced grid while restricting the ``new'' $E(B-V)$ (which is $E(B-V)_\mathrm{SFD} + \delta E(B-V)$) to within -0.05 to 0.25. We also modify the inverse variance of the fluxes such that the S/N is unchanged. We run \texttt{rvspecfit} on the artificially reddened/dereddened spectra, and compare the new synthetic $g-r$ color with the original synthetic color.

\begin{figure}
    \centering
    \includegraphics[width=0.98\columnwidth]{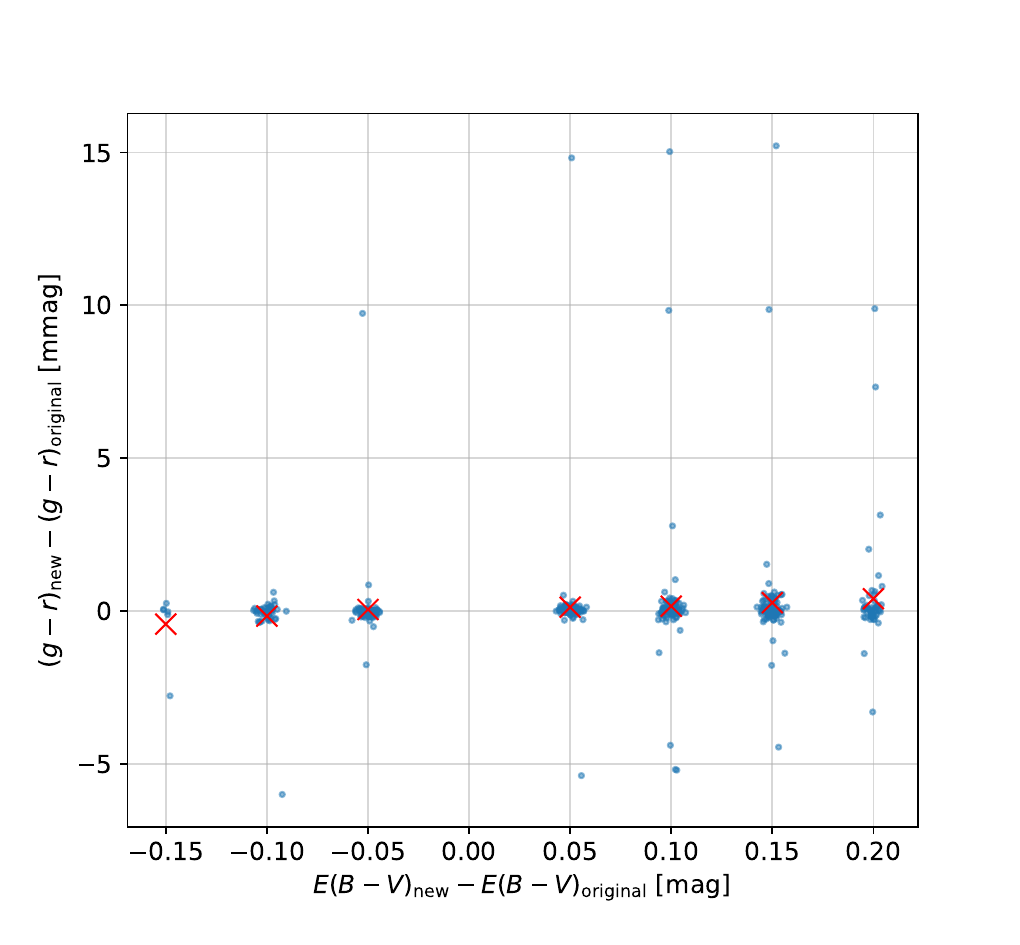}
    \caption{The change in synthetic $g-r$ color as a function of the change in extinction. The blue points are individual stars, and the red crosses show the mean $g-r$ shifts. The largest mean shift is 0.4 mmag. The $E(B-V)$ changes are in 0.05 mag intervals, and they are shifted slightly (along the x-direction) in the figure for better visibility of the data points.}
    \label{fig:extinction_bias}
\end{figure}

The results are shown in Figure \ref{fig:extinction_bias}. We find that the systematic offset, i.e., the mean shift in the synthetic $g-r$ color at each $\delta E(B-V)$, is less than 0.5 mmag. For comparison, the smallest statistical error (due to measurement uncertainties) in a 13.7$'$ (NSIDE=256) pixel is 1.9 mmag, and the systematic error due to photometric calibration uncertainties is roughly 6 mmag. We thus conclude that the dependence of the synthetic colors on the amount of extinction is mostly negligible.

It is possible to imagine that for a very high amount of extinction, the signal-to-noise ratio may be significantly low for the blue part of the spectrum. That may lead to stellar parameters to be predominantly based on the red part of the spectrum which may lead to different stellar parameters (because of the mismatch between models and true spectrum). However, we expect that this requires a much higher level of extinction than the typical extinction within the DESI survey footprint.

\subsection{Dependence of $R_x$ on the stellar spectrum}
\label{sec:rx_stellar_spectrum_dependence}
As we measure the reddening in broad bandpasses, the amount of reddening in the broad-band colors has some dependence on the intrinsic spectrum of the star. We have ignored this dependence in this paper, but we can estimate its magnitude using the \texttt{rvspecfit} synthetic spectra. While we know that the \texttt{rvspecfit} models are not perfect (and we had to correct for systematic offsets in the synthetic colors in Section \ref{sec:calibration}), this exercise nonetheless gives us an idea of the size of this effect. We find that for a fixed $E(B-V)$, the measured reddening increases with $T_\mathrm{eff}$ –– this is expected as a hotter star shifts the effective wavelength of the filter bluer and thus experiences more reddening. The hottest star in our sample, with $T_\mathrm{eff}=6500$ K, has $\sim 4\%$ more reddening in $g-r$ and $\sim 1\%$ more reddening in $r-z$ than the coolest star with 5000 K. This effect is proportional to $E(B-V)$. At the median $E(B-V)$ of 0.05 mag of the DESI footprint, this corresponds to a difference of 2 mmag in $E(g-r)$ and 0.5 mmag in $E(r-z)$.

The size of this effect is much smaller than the random errors in the reddening measurements, and it is further mitigated by the fact that the effect averages out when there are many stars with different $T_\mathrm{eff}$ in each pixel. Therefore we consider this effect to be negligible.

\subsection{Errors in the assumed filter curves}
\label{sec:filter_errors}
The extinction coefficient $R_x$ of each filter is determined based on the assumed filter curve, i.e., the total throughput as a function of wavelength, which depends on the atmospheric transmission, telescope optics, filter transmission and CCD quantum efficiency (QE). Any errors in the filter transmission and CCD QE measurements can lead to a systematic offset in $R_x$. Moreover, while we assume constant filter curves in this work, this is not strictly true as the atmospheric transmission varies with airmass and precipitable water vapor. Since these errors can be quantified as a shift in the effective wavelength, we assess the significance of this effect by slightly shifting the filter curve and remeasuring $R_x$. We find that, if there is a +10 \textup{~\AA} shift for all the filter curves, $R_x$ shifts by $-0.2\%$ in $g$, $r$ and $z$ bands, and $R_g-R_r$ and $R_r-R_z$ shifts by $-0.2\%$ and $-0.3\%$, respectively. Therefore the impact of such as shift is very small ––– it is less than 1 mmag for $E(B-V)<0.3$. The coefficients are only slightly dependent on airmass, with the largest effect in $g$-band where the (DECam) coefficient would be 3.219 at airmass=1 and 3.202 at airmass=2.

\section{Data availability}
\label{sec:data_availability}
The reddening maps are immediately available\footnote{\url{https://data.desi.lbl.gov/public/papers/mws/desi_dust/y2/v1/maps/}}; they are provided for both $g-r$ and $r-z$ colors, with different resolutions (NSIDE from 128 to 512). For convenience, we also provide the $E(B-V)$ values inferred from $g-r$ and $r-z$ via equations \ref{eq:egr_to_ebv} and \ref{eq:erz_to_ebv}. Since the $g-r$ map has higher completeness and is slightly less noisy, we recommend using the $g-r$ map for general purposes. In principle, we could have created a $g-z$ map or a weighted average of the $g-r$ and $r-z$ maps, but we opted not to because of the lower completeness of such a map due to the large number of stars that are saturated in $z$ band (especially in the Northern imaging). Additional data for reproducing figures in this paper is also available\footnote{\url{https://zenodo.org/records/13695452}}.

The first-year portion of the observed spectra, \texttt{rvspecfit} synthetic spectra, and catalogs containing the stellar parameters and synthetic photometry have been released as Value-added Catalogs in DESI Data Release 1 \citep{collaborationDataRelease12025}\footnote{\url{https://data.desi.lbl.gov/doc/releases/dr1/vac/stellar-reddening/}}. The synthetic spectra can be used to measure the reddening using imaging data from other surveys and in different filters (potentially with better photometric calibration). The remainder of the data (taken during the second year of observations and processed in the ``daily'' reduction) will be made public with DESI Data Release 2.

As DESI continues to collect data, we plan to release updated reddening maps in the future with improved coverage and accuracy.

\section{Conclusion}
\label{sec:conclusion}

We have presented Galactic reddening measurements using stellar spectroscopy from the first two years of DESI observations, and we use these measurements to construct new reddening maps that cover approximately 14,000 square degrees. The reddening measurements are made in $g-r$ and $r-z$ colors, with per-star uncertainties of 20 mmag and 22 mmag, respectively.

We carefully remove stars with potentially unreliable measurements and calibrate the synthetic colors to remove systematic offsets arising from inaccuracies in the synthetic spectra. In addition to the reddening measurement itself, we also estimate the error in the measurement by comparing the map-level measurements, accounting for error variations with stellar parameters and spectroscopic S/N. This allows us to optimally combine reddening measurements of individual stars into HEALPix maps.

A comparison between our reddening map and the commonly-used SFD map reveals significant systematic errors in the SFD map. This explains angular patterns in the densities of DESI emission line galaxies that were selected using SFD for extinction correction and whose angular density is up to 80\% correlated with the SFD errors at large scales. The ELG sample re-selected using the DESI reddening map is significantly more uniform. We also find that our reddening map contains smaller-scale variations that are not entirely captured by the SFD map.

We validate the $R_\mathrm{V}=3.1$ \cite{fitzpatrick_correcting_1999} extinction curve by comparing the reddening in $g-r$ and in $r-z$, and we find that on average they are in reasonably good agreements within our survey footprint, thus reaffirming the finding in \cite{schlafly_measuring_2011}. However, the photometric calibration errors, which are about 6 mmag in both colors, make it difficult to interpret the differences between the $g-r$ and $r-z$ measurements. The photometric calibration errors impose a fundamental limit on the accuracy of our reddening maps.
\\

We would like to thank Douglas Finkbeiner for helpful discussions during preparations of the dust map, and Leandro Beraldo e Silva, Carlos Allende Prieto and the anonymous referee for their helpful comments on the draft. R.Z. is supported by the Director, Office of Science, Office of High Energy Physics of the U.S. Department of Energy under Contract No.\ DE–AC02–05CH11231. This material is based upon work supported by the U.S. Department of Energy (DOE), Office of Science, Office of High-Energy Physics, under Contract No. DE–AC02–05CH11231, and by the National Energy Research Scientific Computing Center, a DOE Office of Science User Facility under the same contract. Additional support for DESI was provided by the U.S. National Science Foundation (NSF), Division of Astronomical Sciences under Contract No. AST-0950945 to the NSF’s National Optical-Infrared Astronomy Research Laboratory; the Science and Technology Facilities Council of the United Kingdom; the Gordon and Betty Moore Foundation; the Heising-Simons Foundation; the French Alternative Energies and Atomic Energy Commission (CEA); the National Council of Humanities, Science and Technology of Mexico (CONAHCYT); the Ministry of Science and Innovation of Spain (MICINN), and by the DESI Member Institutions: \url{https://www.desi.lbl.gov/collaborating-institutions}. Any opinions, findings, and conclusions or recommendations expressed in this material are those of the author(s) and do not necessarily reflect the views of the U. S. National Science Foundation, the U. S. Department of Energy, or any of the listed funding agencies.

The authors are honored to be permitted to conduct scientific research on Iolkam Du’ag (Kitt Peak), a mountain with particular significance to the Tohono O’odham Nation.

This work has made use of data from the European Space Agency (ESA) mission {\it Gaia} (\url{https://www.cosmos.esa.int/gaia}), processed by the {\it Gaia} Data Processing and Analysis Consortium (DPAC, \url{https://www.cosmos.esa.int/web/gaia/dpac/consortium}). Funding for the DPAC has been provided by national institutions, in particular the institutions
participating in the {\it Gaia} Multilateral Agreement.

\vspace{5mm}
\textit{Software:} Astropy \citep{astropy:2013, astropy:2018, astropy:2022}, HEALPix/HEALpy \citep{healpix, healpy}, Matplotlib \citep{matplotlib}, Numpy \citep{numpy}, statsmodels \citep{seabold2010statsmodels}, speclite \citep{david_kirkby_2024_13225530}, \texttt{redrock} \citep{bailey_redrock_2024}, \texttt{rvspecfit} \citep{koposov_rvspecfit_2019, koposov_accurate_2011}.

\appendix

\section{Gaia-to-LS transformation}
\label{sec:gaia-to-ls}

The Gaia-to-LS transformation is based on polynomials of the $G_\mathrm{BP}-G_\mathrm{RP}$ color fit to the difference between LS magnitude and Gaia $G$-band magnitude:
\begin{equation}
    m_\mathrm{LS} = G + a_0 + a_1 (G_\mathrm{BP}-G_\mathrm{RP}) + a_2 (G_\mathrm{BP}-G_\mathrm{RP})^2 + ...
\end{equation}
where $m_\mathrm{LS}$ is the LS magnitude (in $g$, $r$ or $z$ band) and $G$, $G_\mathrm{BP}$, $G_\mathrm{RP}$ are the Gaia magnitudes.

\begin{figure*}
    \centering
    \includegraphics[width=2.0\columnwidth]{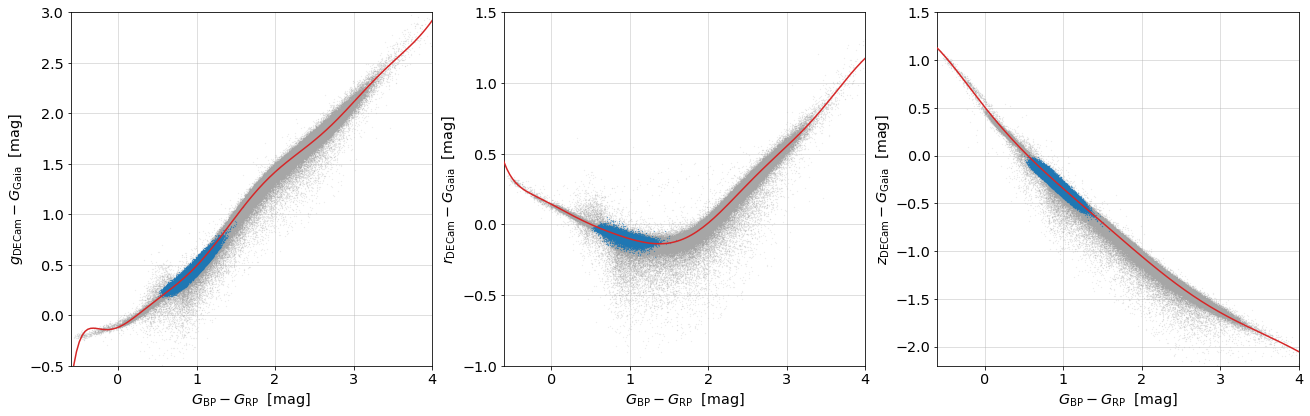}
    \caption{Gaia transformations for DECam $g$, $r$, and $z$ bands. The gray points are stars used for the polynomial fit. The blue points are DESI stars used in the reddening measurements. The red curves are the polynomial fits.}
    \label{fig:gaia_transform}
\end{figure*}

We use stars with $16<G<18$ for the fit, and we apply the imaging quality cuts similar to those in Table \ref{tab:quality_cuts}. The fitting is done separately for the Northern and Southern imaging. We use 12-th order polynomials for the $g$ and $r$ bands and 7-th order for the $z$ band, using the Robust Linear Models (\texttt{RLM}) routine from \texttt{statsmodels}. The scatter, in normalized median absolute deviation, in the predicted LS magnitudes is 18, 11, 17 mmag in $g$, $r$, $z$ bands, respectively (the numbers are similar for the DESI stars used in the reddening measurements). The Gaia-predicted LS magnitudes have been used in the bright star ``halo'' subtraction in the Legacy Surveys pipeline since DR9\footnote{\url{https://www.legacysurvey.org/dr9/psf/}}, and the high polynomials are necessary to accurately predict the LS magnitudes of the full range of stars (without color cuts). The DESI stars used for reddening only occupy a narrow range in $G_\mathrm{BP}-G_\mathrm{RP}$ color. Figure \ref{fig:gaia_transform} shows the transformations to the DECam filters.

\section{Filling in missing pixels}
\label{sec:filling}

Here we describe the procedure to fill in (or ``inpaint'') pixels with no stars via interpolation. We use the \texttt{sphtfunc.smoothing} routine in \texttt{healpy} for the Gaussian smoothing. Because the routine does not distinguish between masked pixels in the convolution, we initialize the map by filling in all missing pixels with the median $E(a-b)$ value of the valid pixels. We then convolve the map with a Gaussian kernel with FWHM of 0.5$^{\circ}$/0.25$^{\circ}$/0.125$^{\circ}$ for NSIDE of 128/256/512, but only updating the missing pixels and keeping the valid pixels unchanged. After the convolution, the inpainted pixels will be biased towards the initial fill value, and we iterate this process to reduce this bias. We iterate 200 to 800 times (depending on the resolution) such that it converges for missing pixels inside the survey footprint.

Inpainted pixels that are farther away from valid pixels are more likely to have catastrophically wrong estimates than pixels that are closer to valid pixels. To distinguish between them, we generate a separate ``fill fraction'' map tracking the contribution of valid pixels. The procedure is identical to that of the $E(a-b)$ map, but replacing the $E(a-b)$ value with 0 for valid pixels and 1 for missing pixels. We consider a pixel to have a ``good'' interpolated value if the fill fraction is less than 0.01/0.005/0.001 for NSIDE of 128/256/512. This ``good interpolation'' definition is used for the $C_\ell$ calculations in which the qualified interpolated pixels are counted as observed pixels for the sky fraction.

\bibliography{DESI_dust}{}

\begin{thebibliography}{}
\makeatletter
\relax
\def\mn@urlcharsother{\let\do\@makeother \do\$\do\&\do\#\do\^\do\_\do\%\do\~}
\def\mn@doi{\begingroup\mn@urlcharsother \@ifnextchar [ {\mn@doi@} {\mn@doi@[]}}
\def\mn@doi@[#1]#2{\def\@tempa{#1}\ifx\@tempa\@empty \href {http://dx.doi.org/#2} {doi:#2}\else \href {http://dx.doi.org/#2} {#1}\fi \endgroup}
\def\mn@eprint#1#2{\mn@eprint@#1:#2::\@nil}
\def\mn@eprint@arXiv#1{\href {http://arxiv.org/abs/#1} {{\tt arXiv:#1}}}
\def\mn@eprint@dblp#1{\href {http://dblp.uni-trier.de/rec/bibtex/#1.xml} {dblp:#1}}
\def\mn@eprint@#1:#2:#3:#4\@nil{\def\@tempa {#1}\def\@tempb {#2}\def\@tempc {#3}\ifx \@tempc \@empty \let \@tempc \@tempb \let \@tempb \@tempa \fi \ifx \@tempb \@empty \def\@tempb {arXiv}\fi \@ifundefined {mn@eprint@\@tempb}{\@tempb:\@tempc}{\expandafter \expandafter \csname mn@eprint@\@tempb\endcsname \expandafter{\@tempc}}}

\bibitem[\protect\citeauthoryear{Alberto et~al.}{Alberto et~al.}{2024}]{alberto_elg_imaging_sys}
Alberto R.-M.,  et~al., 2024, in preparation

\bibitem[\protect\citeauthoryear{{Astropy Collaboration} et~al.,}{{Astropy Collaboration} et~al.}{2013}]{astropy:2013}
{Astropy Collaboration} et~al., 2013, \mn@doi [\aap] {10.1051/0004-6361/201322068}, \href {http://adsabs.harvard.edu/abs/2013A%26A...558A..33A} {558, A33}

\bibitem[\protect\citeauthoryear{{Astropy Collaboration} et~al.,}{{Astropy Collaboration} et~al.}{2018}]{astropy:2018}
{Astropy Collaboration} et~al., 2018, \mn@doi [\aj] {10.3847/1538-3881/aabc4f}, \href {https://ui.adsabs.harvard.edu/abs/2018AJ....156..123A} {156, 123}

\bibitem[\protect\citeauthoryear{{Astropy Collaboration} et~al.,}{{Astropy Collaboration} et~al.}{2022}]{astropy:2022}
{Astropy Collaboration} et~al., 2022, \mn@doi [\apj] {10.3847/1538-4357/ac7c74}, \href {https://ui.adsabs.harvard.edu/abs/2022ApJ...935..167A} {935, 167}

\bibitem[\protect\citeauthoryear{Bailey et~al.}{Bailey et~al.}{2025}]{bailey_redrock_2024}
Bailey S.,  et~al., 2025, in preparation

\bibitem[\protect\citeauthoryear{Burstein \& Heiles}{Burstein \& Heiles}{1982}]{bursteinReddeningsDerivedGalaxy1982}
Burstein D.,  Heiles C.,  1982, \mn@doi [The Astronomical Journal] {10.1086/113199}, 87, 1165

\bibitem[\protect\citeauthoryear{Chambers et~al.,}{Chambers et~al.}{2019}]{chambers_panstarrs1_2019}
Chambers K.~C.,  et~al., 2019, The {{Pan-STARRS1 Surveys}} (\mn@eprint {arXiv} {1612.05560})

\bibitem[\protect\citeauthoryear{Chiang}{Chiang}{2023}]{chiang_corrected_2023}
Chiang Y.-K.,  2023, Corrected {{SFD}}: {{A More Accurate Galactic Dust Map}} with {{Minimal Extragalactic Contamination}} (\mn@eprint {arXiv} {2306.03926})

\bibitem[\protect\citeauthoryear{Chiang \& M{\'e}nard}{Chiang \& M{\'e}nard}{2019}]{chiang_extragalactic_2019}
Chiang Y.-K.,  M{\'e}nard B.,  2019, \mn@doi [The Astrophysical Journal] {10.3847/1538-4357/aaf4f6}, 870, 120

\bibitem[\protect\citeauthoryear{Cooper et~al.,}{Cooper et~al.}{2023}]{cooperOverviewDESIMilky2023}
Cooper A.~P.,  et~al., 2023, \mn@doi [The Astrophysical Journal] {10.3847/1538-4357/acb3c0}, 947, 37

\bibitem[\protect\citeauthoryear{{DESI Collaboration} et~al.,}{{DESI Collaboration} et~al.}{2016a}]{DESI2016a.Science}
{DESI Collaboration} et~al., 2016a, arXiv e-prints, \href {https://ui.adsabs.harvard.edu/abs/2016arXiv161100036D} { arXiv:1611.00036}

\bibitem[\protect\citeauthoryear{{DESI Collaboration} et~al.,}{{DESI Collaboration} et~al.}{2016b}]{DESI2016b.Instr}
{DESI Collaboration} et~al., 2016b, arXiv e-prints, \href {https://ui.adsabs.harvard.edu/abs/2016arXiv161100037D} { arXiv:1611.00037}

\bibitem[\protect\citeauthoryear{{DESI Collaboration} et~al.,}{{DESI Collaboration} et~al.}{2024}]{DESI_EDR}
{DESI Collaboration} et~al., 2024, \mn@doi [\aj] {10.3847/1538-3881/ad3217}, \href {https://ui.adsabs.harvard.edu/abs/2024AJ....168...58D} {168, 58}

\bibitem[\protect\citeauthoryear{{DESI Collaboration} et~al.,}{{DESI Collaboration} et~al.}{2025}]{collaborationDataRelease12025}
{DESI Collaboration} et~al., 2025, Data {{Release}} 1 of the {{Dark Energy Spectroscopic Instrument}} (\mn@eprint {arXiv} {2503.14745}), \mn@doi{10.48550/arXiv.2503.14745}

\bibitem[\protect\citeauthoryear{{Dark Energy Survey Collaboration} et~al.,}{{Dark Energy Survey Collaboration} et~al.}{2016}]{darkenergysurveycollaboration_dark_2016}
{Dark Energy Survey Collaboration} et~al., 2016, \mn@doi [Monthly Notices of the Royal Astronomical Society] {10.1093/mnras/stw641}, 460, 1270

\bibitem[\protect\citeauthoryear{Dey et~al.,}{Dey et~al.}{2019}]{dey_overview_2019}
Dey A.,  et~al., 2019, \mn@doi [The Astronomical Journal] {10.3847/1538-3881/ab089d}, 157, 168

\bibitem[\protect\citeauthoryear{Dey et~al.,}{Dey et~al.}{2025}]{deyBackupProgramDark2025}
Dey A.,  et~al., 2025, The {{Backup Program}} of the {{Dark Energy Spectroscopic Instrument}}'s {{Milky Way Survey}} (\mn@eprint {arXiv} {2505.17230}), \mn@doi{10.48550/arXiv.2505.17230}

\bibitem[\protect\citeauthoryear{Fitzpatrick}{Fitzpatrick}{1999}]{fitzpatrick_correcting_1999}
Fitzpatrick E.~L.,  1999, \mn@doi [Publications of the Astronomical Society of the Pacific] {10.1086/316293}, 111, 63

\bibitem[\protect\citeauthoryear{{G{\'o}rski}, {Hivon}, {Banday}, {Wandelt}, {Hansen}, {Reinecke}  \& {Bartelmann}}{{G{\'o}rski} et~al.}{2005}]{healpix}
{G{\'o}rski} K.~M.,  {Hivon} E.,  {Banday} A.~J.,  {Wandelt} B.~D.,  {Hansen} F.~K.,  {Reinecke} M.,   {Bartelmann} M.,  2005, \mn@doi [\apj] {10.1086/427976}, \href {http://adsabs.harvard.edu/abs/2005ApJ...622..759G} {622, 759}

\bibitem[\protect\citeauthoryear{Guy et~al.,}{Guy et~al.}{2023}]{guy_spectroscopic_2023}
Guy J.,  et~al., 2023, \mn@doi [The Astronomical Journal] {10.3847/1538-3881/acb212}, 165, 144

\bibitem[\protect\citeauthoryear{Harris et~al.,}{Harris et~al.}{2020}]{numpy}
Harris C.~R.,  et~al., 2020, \mn@doi [Nature] {10.1038/s41586-020-2649-2}, 585, 357

\bibitem[\protect\citeauthoryear{Hunter}{Hunter}{2007}]{matplotlib}
Hunter J.~D.,  2007, \mn@doi [Computing in Science \& Engineering] {10.1109/MCSE.2007.55}, 9, 90

\bibitem[\protect\citeauthoryear{Husser, {von Berg}, Dreizler, Homeier, Reiners, Barman  \& Hauschildt}{Husser et~al.}{2013}]{husser_new_2013}
Husser T.-O.,  {von Berg} S.~W.,  Dreizler S.,  Homeier D.,  Reiners A.,  Barman T.,   Hauschildt P.~H.,  2013, \mn@doi [Astronomy \& Astrophysics] {10.1051/0004-6361/201219058}, 553, A6

\bibitem[\protect\citeauthoryear{Huterer, Cunha  \& Fang}{Huterer et~al.}{2013}]{huterer_calibration_2013}
Huterer D.,  Cunha C.~E.,   Fang W.,  2013, \mn@doi [Monthly Notices of the Royal Astronomical Society] {10.1093/mnras/stt653}, 432, 2945

\bibitem[\protect\citeauthoryear{Kirkby et~al.,}{Kirkby et~al.}{2024}]{david_kirkby_2024_13225530}
Kirkby D.,  et~al., 2024, desihub/speclite: Bug fix release: General clean- up prior to refactoring package infrastructure, \mn@doi{10.5281/zenodo.13225530}, \url {https://doi.org/10.5281/zenodo.13225530}

\bibitem[\protect\citeauthoryear{Koposov}{Koposov}{2019}]{koposov_rvspecfit_2019}
Koposov S.~E.,  2019, Astrophysics Source Code Library,  ascl:1907.013

\bibitem[\protect\citeauthoryear{Koposov et~al.,}{Koposov et~al.}{2011}]{koposov_accurate_2011}
Koposov S.~E.,  et~al., 2011, \mn@doi [The Astrophysical Journal] {10.1088/0004-637X/736/2/146}, 736, 146

\bibitem[\protect\citeauthoryear{{Koposov} et~al.,}{{Koposov} et~al.}{2024}]{desi_edr_mws_vac}
{Koposov} S.~E.,  et~al., 2024, \mn@doi [\mnras] {10.1093/mnras/stae1842}, 533, 1012

\bibitem[\protect\citeauthoryear{{Lang}, {Hogg}  \& {Mykytyn}}{{Lang} et~al.}{2016}]{Lang16}
{Lang} D.,  {Hogg} D.~W.,   {Mykytyn} D.,  2016, {The Tractor: Probabilistic astronomical source detection and measurement}, Astrophysics Source Code Library (\mn@eprint {ascl} {1604.008})

\bibitem[\protect\citeauthoryear{Lenz, Hensley  \& Dor{\'e}}{Lenz et~al.}{2017}]{lenz_new_2017}
Lenz D.,  Hensley B.~S.,   Dor{\'e} O.,  2017, \mn@doi [The Astrophysical Journal] {10.3847/1538-4357/aa84af}, 846, 38

\bibitem[\protect\citeauthoryear{{Manser} et~al.,}{{Manser} et~al.}{2024}]{manser_desi_2024}
{Manser} C.~J.,  et~al., 2024, arXiv e-prints, \href {https://ui.adsabs.harvard.edu/abs/2024arXiv240218641M} { arXiv:2402.18641}

\bibitem[\protect\citeauthoryear{Myers et~al.,}{Myers et~al.}{2023}]{myers_targetselection_2023}
Myers A.~D.,  et~al., 2023, \mn@doi [The Astronomical Journal] {10.3847/1538-3881/aca5f9}, 165, 50

\bibitem[\protect\citeauthoryear{Peek \& Graves}{Peek \& Graves}{2010}]{peek_correction_2010}
Peek J. E.~G.,  Graves G.~J.,  2010, \mn@doi [The Astrophysical Journal] {10.1088/0004-637X/719/1/415}, 719, 415

\bibitem[\protect\citeauthoryear{Raichoor et~al.,}{Raichoor et~al.}{2023}]{raichoor_target_2023}
Raichoor A.,  et~al., 2023, \mn@doi [The Astronomical Journal] {10.3847/1538-3881/acb213}, 165, 126

\bibitem[\protect\citeauthoryear{Ross et~al.,}{Ross et~al.}{2012}]{ross_clustering_2012}
Ross A.~J.,  et~al., 2012, \mn@doi [Monthly Notices of the Royal Astronomical Society] {10.1111/j.1365-2966.2012.21235.x}, 424, 564

\bibitem[\protect\citeauthoryear{Schlafly \& Finkbeiner}{Schlafly \& Finkbeiner}{2011}]{schlafly_measuring_2011}
Schlafly E.~F.,  Finkbeiner D.~P.,  2011, \mn@doi [The Astrophysical Journal] {10.1088/0004-637X/737/2/103}, 737, 103

\bibitem[\protect\citeauthoryear{Schlafly et~al.,}{Schlafly et~al.}{2016}]{schlaflyOpticalinfraredExtinctionCurve2016}
Schlafly E.~F.,  et~al., 2016, \mn@doi [The Astrophysical Journal] {10.3847/0004-637X/821/2/78}, 821, 78

\bibitem[\protect\citeauthoryear{Schlafly et~al.,}{Schlafly et~al.}{2023}]{schlafly_survey_2023}
Schlafly E.~F.,  et~al., 2023, \mn@doi [The Astronomical Journal] {10.3847/1538-3881/ad0832}, 166, 259

\bibitem[\protect\citeauthoryear{Schlegel, Finkbeiner  \& Davis}{Schlegel et~al.}{1998}]{schlegel_maps_1998}
Schlegel D.~J.,  Finkbeiner D.~P.,   Davis M.,  1998, \mn@doi [The Astrophysical Journal] {10.1086/305772}, 500, 525

\bibitem[\protect\citeauthoryear{Schlegel et~al.}{Schlegel et~al.}{2024}]{schlegel_ls_dr9}
Schlegel D.~J.,  et~al., 2024, in preparation

\bibitem[\protect\citeauthoryear{Seabold \& Perktold}{Seabold \& Perktold}{2010}]{seabold2010statsmodels}
Seabold S.,  Perktold J.,  2010, in 9th Python in Science Conference.

\bibitem[\protect\citeauthoryear{{The LSST Dark Energy Science Collaboration} et~al.,}{{The LSST Dark Energy Science Collaboration} et~al.}{2021}]{thelsstdarkenergysciencecollaboration_lsst_2021}
{The LSST Dark Energy Science Collaboration} et~al., 2021, The {{LSST Dark Energy Science Collaboration}} ({{DESC}}) {{Science Requirements Document}} (\mn@eprint {arXiv} {1809.01669})

\bibitem[\protect\citeauthoryear{Zhang \& Green}{Zhang \& Green}{2025}]{zhangThreedimensionalMapsInterstellar2025}
Zhang X.,  Green G.~M.,  2025, \mn@doi [Science] {10.1126/science.ado9787}, 387, 1209

\bibitem[\protect\citeauthoryear{Zhou, Dey, Lang, Moustakas, Schlafly  \& Schlegel}{Zhou et~al.}{2023a}]{zhou_comparing_2023}
Zhou R.,  Dey A.,  Lang D.,  Moustakas J.,  Schlafly E.~F.,   Schlegel D.~J.,  2023a, \mn@doi [Research Notes of the AAS] {10.3847/2515-5172/acd7ef}, 7, 105

\bibitem[\protect\citeauthoryear{Zhou et~al.,}{Zhou et~al.}{2023b}]{zhou_target_2023}
Zhou R.,  et~al., 2023b, \mn@doi [The Astronomical Journal] {10.3847/1538-3881/aca5fb}, 165, 58

\bibitem[\protect\citeauthoryear{Zonca, Singer, Lenz, Reinecke, Rosset, Hivon  \& Gorski}{Zonca et~al.}{2019}]{healpy}
Zonca A.,  Singer L.,  Lenz D.,  Reinecke M.,  Rosset C.,  Hivon E.,   Gorski K.,  2019, \mn@doi [Journal of Open Source Software] {10.21105/joss.01298}, 4, 1298

\bibitem[\protect\citeauthoryear{Zou et~al.,}{Zou et~al.}{2017}]{zou_project_2017}
Zou H.,  et~al., 2017, \mn@doi [Publications of the Astronomical Society of the Pacific] {10.1088/1538-3873/aa65ba}, 129, 064101

\makeatother
\end{thebibliography}
\bibliographystyle{mnras}

\end{document}